\documentclass[pre,twocolumn,showpacs,preprintnumbers,amsmath,amssymb]{revtex4}

\usepackage{graphicx}
\usepackage{bm}
\usepackage{hyperref}
\hypersetup{colorlinks = true, urlcolor = blue, linkcolor = blue, citecolor = blue}
\usepackage{color}
\usepackage{ulem}

\begin{document}


\title{Boson peak and Ioffe-Regel criterion in amorphous silicon-like materials:\\ the effect of bond directionality}

\author{Y. M. Beltukov}
\affiliation{Ioffe Physical Technical Institute, 194021 St Petersburg, Russian Federation}
\affiliation{Universit\'e Montpellier II, CNRS, Montpellier 34095, France}
\author{C. Fusco}
\affiliation{Universit\'e de Lyon, MATEIS, INSA-Lyon, CNRS UMR5510, F-69621, France}
\affiliation{Institut Lumi\`ere  Mati\`ere, UMR 5306 Universit\'e Lyon 1-CNRS,  F-69622 Villeurbanne Cedex, France}
\author{D. A. Parshin}
\affiliation{Saint Petersburg State Polytechnical University, 195251 Saint Petersburg, Russian Federation}
\author{A. Tanguy}
\affiliation{Universit\'e de Lyon, LaMCoS, INSA-Lyon, CNRS UMR5259, F-69621, France}
\affiliation{Institut Lumi\`ere  Mati\`ere, UMR 5306 Universit\'e Lyon 1-CNRS,  F-69622 Villeurbanne Cedex, France}

\begin{abstract}
The vibrational properties of model amorphous materials are studied by combining complete analysis of the vibration modes, dynamical structure factor and energy diffusivity with exact diagonalization of the dynamical matrix and the Kernel Polynomial Method which allows a study of very large system sizes. Different materials are studied that differ only by the bending rigidity of the interactions in a Stillinger-Weber modelization used to describe amorphous silicon. The local bending rigidity can thus be used as a control parameter, to tune the sound velocity together with local bonds directionality. It is shown that for all the systems studied, the upper limit of the Boson peak corresponds to the Ioffe-Regel criterion for transverse waves, as well as to a minimum of the diffusivity. The Boson peak is followed by a diffusivity's increase supported by longitudinal phonons. The Ioffe-Regel criterion for transverse waves corresponds to a common characteristic mean-free path of 5--7 \AA{} (which is slightly bigger for longitudinal phonons), while the fine structure of the vibrational density of states is shown to be sensitive to the local bending rigidity.
\end{abstract}


\maketitle

\section{Introduction}
\label{sec:intro}

Amorphous silicon is a model material studied for a long time as a simple example of monoatomic amorphous material with important possible applications to electronics (as a semi-conductor device) as well as for photo-voltaic devices~\cite{aSi-Appli}. It is thus important to understand, and if possible control, heat conduction and energy diffusivity in this material. Allen and Feldman studied its vibrational properties a long time ago~\cite{b.feldman0,b.feldman1,b.allen}. They have shown that the majority of the vibrations are not plane waves: propagative modes, including plane waves, being restricted to the low frequency domain, and localized modes occupying the high frequency tail of the spectrum. From an experimental point of view, amorphous silicon is an ubiquitous disordered solid, and the existence of a glass transition  is still a matter of debate in this case~\cite{Tg-aSi2004,Tg-aSi2003,Tg-aSi2004b}. Another question concerns the existence of an excess of low-frequency vibrations as compared to the Debye prediction, also called the Boson peak and widely studied in amorphous materials~\cite{Buchenau1984,Sokolov1986,Sokolov1992,Parshin1992,Parshin1993,Parshin1994,Parshin2003,Buchenau1986,b.schirmacher1,b.schirmacher2,Wyart2005,giordano_2010,giordano_2011,Ruocco2013}. The Boson peak is related to an increase and an anomalous T-dependence of the heat capacity of amorphous samples as compared to crystalline samples~\cite{Pohl1971,b.malinovsky,Parshin2007-Review,Wyart2010b}. The origin of the Boson peak is still debated in amorphous silicon samples~\cite{b.finkemeier,b.resp-finkemeier,He,b.fabian2,aSi-VDOS-Exp,Sokolov1991,Thorpe1973} and a common interpretation of the detailed numerical analysis of its vibrational properties is still lacking~\cite{b.christie,b.marinov,b.fabian2}. However, pure amorphous silicon can also be considered as a simple paragon for amorphous materials, with a monoatomic piling and local tetrahedral order. We will consider here a Stillinger-Weber modelization of amorphous silicon as a model amorphous material used to understand in a systematic way the effect of the local bending rigidity, or bond's directionality, on the vibrational properties at different frequencies.

In amorphous materials, vibrations differ from the usual description in terms of transverse and longitudinal acoustic plane waves and optic modes valid in crystalline materials~\cite{Kittel-book,Phillips-book, taraskin-1997,Duval2007}. More precisely, due to structural disorder, plane waves (also usually referred to as phonons, with a well defined wave vector) are not the vibrational eigenmodes in amorphous materials, contrary to crystalline materials. It results in an apparent amplitude's decay, or progressive scattering, when they propagate through the samples~\cite{taraskin2000,b.tanguy2010}. Moreover, the proximity to small energy barriers gives rise to low frequency soft modes with local quadrupolar shape~\cite{Lemaitre2004,Lemaitre2006,b.tanguy20061} that can be used as predictor for plastic deformation~\cite{b.tanguy2010} and affects the low frequency part of the vibrational response~\cite{Wyart2010b}. At higher frequencies, strong scattering gives rise to a diffusive propagation of vibrational energy. Allen and Feldman have proposed to call ``diffusons'' the eigenmodes in this regime~\cite{b.feldman1}. ``Diffusons'' are extended modes that are different from plane waves and give rise to strong scattering and diffusive propagation of vibrational energy. The crossover from weak to strong scattering is usually determined by the Ioffe-Regel criterion~\cite{Ioffe} that compares the inverse wave vector of the propagative phonon to its mean free path. The connexion between Ioffe-Regel criterion and Boson peak was questioned in previous studies~\cite{taraskin2000b,schober2004,Beltukov2013}. At even higher frequencies, the comparison of the vibrational density of states of the amorphous material with that of the crystal shows a reminiscence of the crystals optic modes~\cite{Sokolov1991,Damart2015} and the existence of a mobility edge~\cite{b.allen,taraskin2002}. Despite important numerical effort in the last years to describe the vibrational response of amorphous materials~\cite{larkin,b.tanguy2002,b.fabian2,b.christie,Wyart2010,Barrat2014}, many important questions are still open, such as the identification of the frequency separating the different regimes, the description of the scattering processes, and the connection with quasi monochromatic wave packets propagation (diffusion of energy). A promising way was proposed with the study of vibrational eigenmodes as eigenvectors of a random matrix with specific properties~\cite{b.taraskin,Beltukov2013}. We will follow the same approach as detailed in~\cite{Beltukov2013} but with a dynamical matrix resulting from molecular dynamics simulations of large-scale amorphous silicon-like systems at equilibrium. We will thus focus on the harmonic contribution of the disorder to the vibrational response of our atomistic model.

The numerical study of the harmonic vibrational properties of amorphous materials involves the dynamical matrix whose inverse is simply the linear approximation of the Green's function relating the displacements to the forces at equilibrium. One possible empirical description of the interactions in amorphous silicon-like materials is due to Stillinger and Weber~\cite{b.stillinger} and involves a simple additive decomposition of the interatomic interactions into two-body interactions (depending only on the distance between atoms) and three-body interaction (involving the relative angle between adjacent bonds). This last term is a way to introduce the degree of covalent bonding. It was shown that the amplitude of this term can affect strongly the mechanical response, such as the yield stress~\cite{b.fusco2010}, the viscosity~\cite{b.fusco2014} or the toughness of the materials~\cite{b.pizza2013}. In this paper, we will use this parameter to tune the sound waves velocity and to show its effect on the anomalous vibrational properties of the systems. We devoted a special attention to the effect of the three-body interactions on the characteristic length scales such as the Boson peak's wavelength, and  the mean-free paths.

The paper is organized as follows: after a detailed presentation of the numerical model, we will first compute the vibrational density of states (Sec.~\ref{sec:DOS}) together with a detailed analysis of the corresponding eigenmodes (Sec.~\ref{sec:modes}). We will then study the dynamical structure factor, the dispersion laws, the corresponding phonon lifetimes and mean-free paths (Sec.~\ref{sec:sf}). Finally, we will compare the results to the propagation of quasi monochromatic wave packets with different frequencies allowing to measure the diffusivity of vibration energy in the materials, as a function of the frequency and of the bending rigidity (Sec.~\ref{sec:diff}). This will allow to identify coherently well defined vibrational domains, as it will be summarized in the conclusion (Sec.~\ref{sec:Conclusion}).

\section{Numerical model\label{sec:num}}

We have studied the vibrational properties of a model amorphous silicon (a-Si) system consisting of $N=32768$ atoms contained in a cubic box of lengths $L_x=L_y=L_z$ of approximately 87 \AA{} (smaller systems of $N=8000$ have also been studied to compare our results). The technical details of the preparation of the a-Si model have already been presented in Ref.~\cite{b.fusco2010}.
The Si-Si interaction in the system studied here is described by the Stillinger-Weber potential~\cite{b.stillinger}, where we have tuned the prefactor $\Lambda$ of the three-body term as already done in previous work~\cite{b.fusco2010,b.fusco2014}, to quantify here the effect of local order on the vibrational properties.
The Stillinger-Weber potential is an empirical potential including two-body and three-body interactions, such that the total potential energy of the system is written as
\begin{multline}
    U=\sum_{i<j}f(r_{ij}) + \Lambda\sum_{i<j<k}\Bigl(g(r_{ij},r_{ik},\theta_{jik})\\+g(r_{ji},r_{jk},\theta_{ijk})+g(r_{ki},r_{kj},\theta_{ikj})\Bigr)
\end{multline}
with
\begin{equation}
    f(r_{ij})=A\left(B/r^4-1\right)\exp{\left(\sigma\left(r_{ij}-a\right)^{-1}\right)}
\end{equation}
and
\begin{multline}
    g(r_{ij},r_{ik},\theta_{jik}) = \left(\cos\theta_{jik}+1/3\right)^2\\
    \times\exp{\left(\alpha (r_{ij}-a)^{-1}+\alpha (r_{ik}-a)^{-1}\right)}\nonumber
\end{multline}
with the parameters proposed in~\cite{b.stillinger} $A=7.05$, $B=11.60$, $\alpha=2.51$ \AA, $\sigma=2.06$ \AA, and $a=3.77$ \AA.
The parameter $\Lambda$ gives a measure of the bond's directionality: high values of $\Lambda$ favor local tetragonal order ($\Lambda=21$ is the original value proposed by Stillinger et al~\cite{b.stillinger} as an empirical model for a-Si). The atomic configurations corresponding to a-Si structures for different values of $\Lambda$ have been obtained from the liquid state, using the open source LAMMPS package~\cite{b.lammps} for classical Molecular Dynamics simulations, and following the procedure already described in~\cite{b.fusco2010}. Different configurations have been obtained, either with a quenching in the NVT ensemble at a fixed density $\rho=2.303$ g/cm${}^3$, giving rise to different final pressures as detailed in Table~\ref{Table1}, either after pressure relaxation up to $P\approx 0$ GPa.

\begin{table}[b]
    \caption{Transverse and longitudinal sound velocities obtained from the elastic moduli for different values of the parameter $\Lambda$ with $N=32 768$.}
    \begin{tabular}{*{6}{@{\hspace{0.2cm}}c@{\hspace{0.2cm}}}}
        \hline\hline
        $\Lambda$ & $\rho$, g/cm${}^3$ & $P$, GPa & $c_T$, m/s & $c_L$, m/s \\
        \hline
        17   & 2.303 & $-1.82$ & 3334 & 7833 \\
        19    & '' & $-0.096$  & 3570 & 7750 \\
        21    & '' &  0.638    & 3854 & 7965 \\
        23.5  & '' &  1.38     & 4133 & 8226 \\
        26.25 & '' &  2.1      & 4386 & 8484 \\
        40    & '' &  5.07     & 5305 & 9490 \\
        \hline
        17  & 2.339 & $-0.011$ & 3312 & 8436 \\
        21  & 2.295 & $0.013$  & 3714 & 8367 \\
        40  & 2.248 & $-0.114$ & 5118 & 9350 \\
        \hline\hline
    \end{tabular}
    \label{Table1}
\end{table}

In order to study the role of the local order on the vibrational properties of a-Si we have calculated the dynamical matrices for different values of $\Lambda$. The dynamical matrix has been numerically computed by calculating the second order spatial derivative of the potential energy around the equilibrium atomic positions ${\bf R}_i$:
\begin{equation}
    M_{i\alpha,j\beta} = \frac{1}{\sqrt{m_i m_j}}\frac{\partial^2 U}{\partial r_{i\alpha}\partial r_{j\beta}}.
\end{equation}
The Newton's equations of motion can thus be written in the harmonic approximation
\begin{equation}
    \ddot{u}_{i\alpha}= -\sum_{j\beta}M_{i\alpha,j\beta}u_{j\beta}   \label{eq:motion}
\end{equation}
with ${\bf u}_i=\sqrt{m_i}({\bf r}_i-{\bf R}_i)$ where ${\bf r}_i-{\bf R}_i$ is the displacement of atom $i$, ${\bf R}_i$ its equilibrium position, $m_i$ its mass, and greek indices $\alpha$ or $\beta$ indicate spatial directions. The elastic constants (shear and bulk modulus) are obtained as in~\cite{b.fusco2010} by measuring the quasi-static response of the system to a small deformation of the box. The corresponding values of the transverse $c_T$ and longitudinal $c_L$ sound velocities are summarized  in Table \ref{Table1} for the different values of $\Lambda$.

\section{Density of states\label{sec:DOS}}

\begin{figure}[t]
    \includegraphics[scale=0.4]{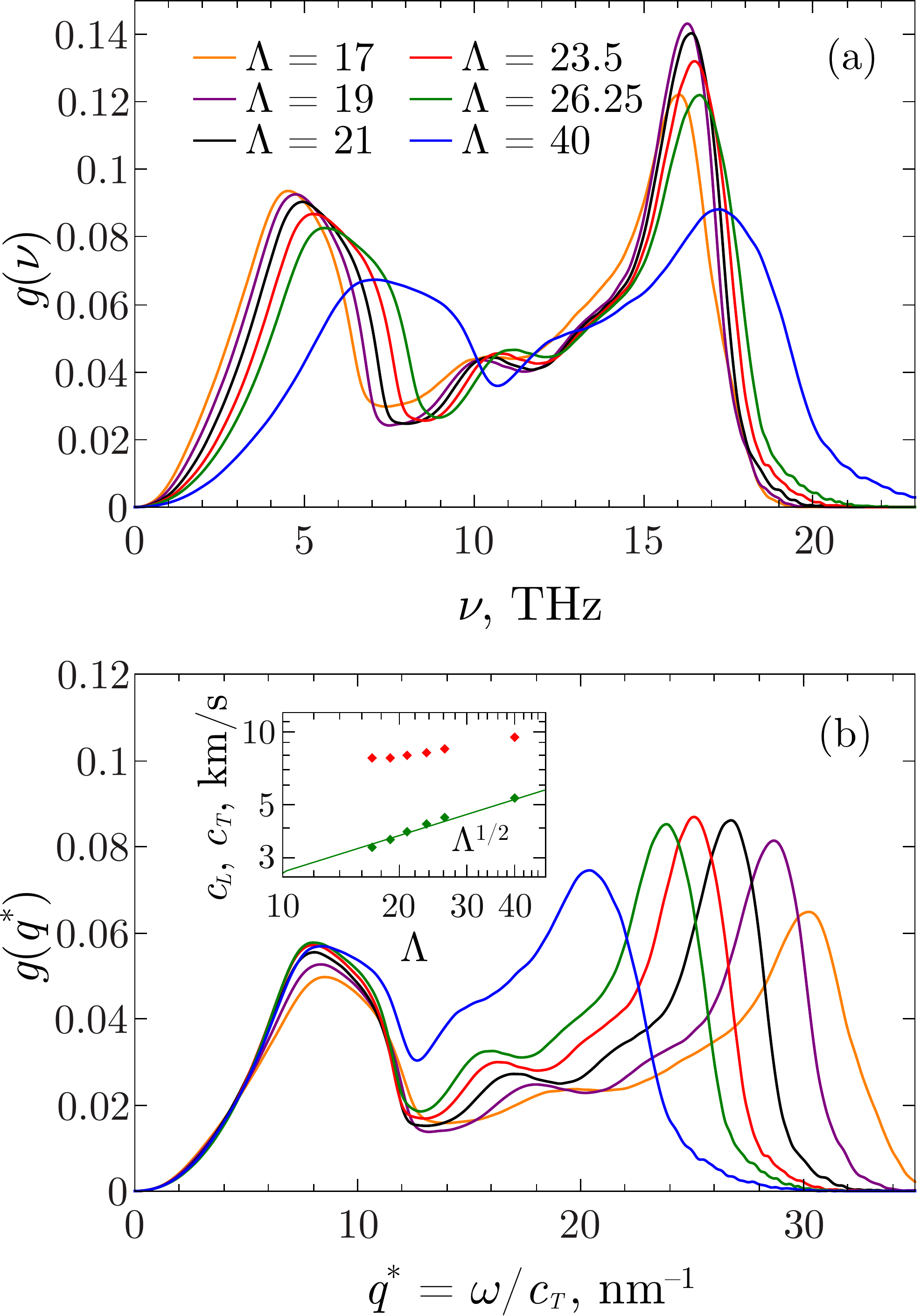}
    \caption{(a) The VDOS for different values of the parameter $\Lambda$. (b) The VDOS as a function of the reduced wave vector $q^* = \omega/c_T$. Inset: $\Lambda$-dependence of the sound velocities $c_T$ and $c_L$. Line shows fit with $c_T \propto \sqrt{\Lambda}$ dependence.}
    \label{f.DOS}
\end{figure}

\begin{figure}
    \includegraphics[scale=0.4]{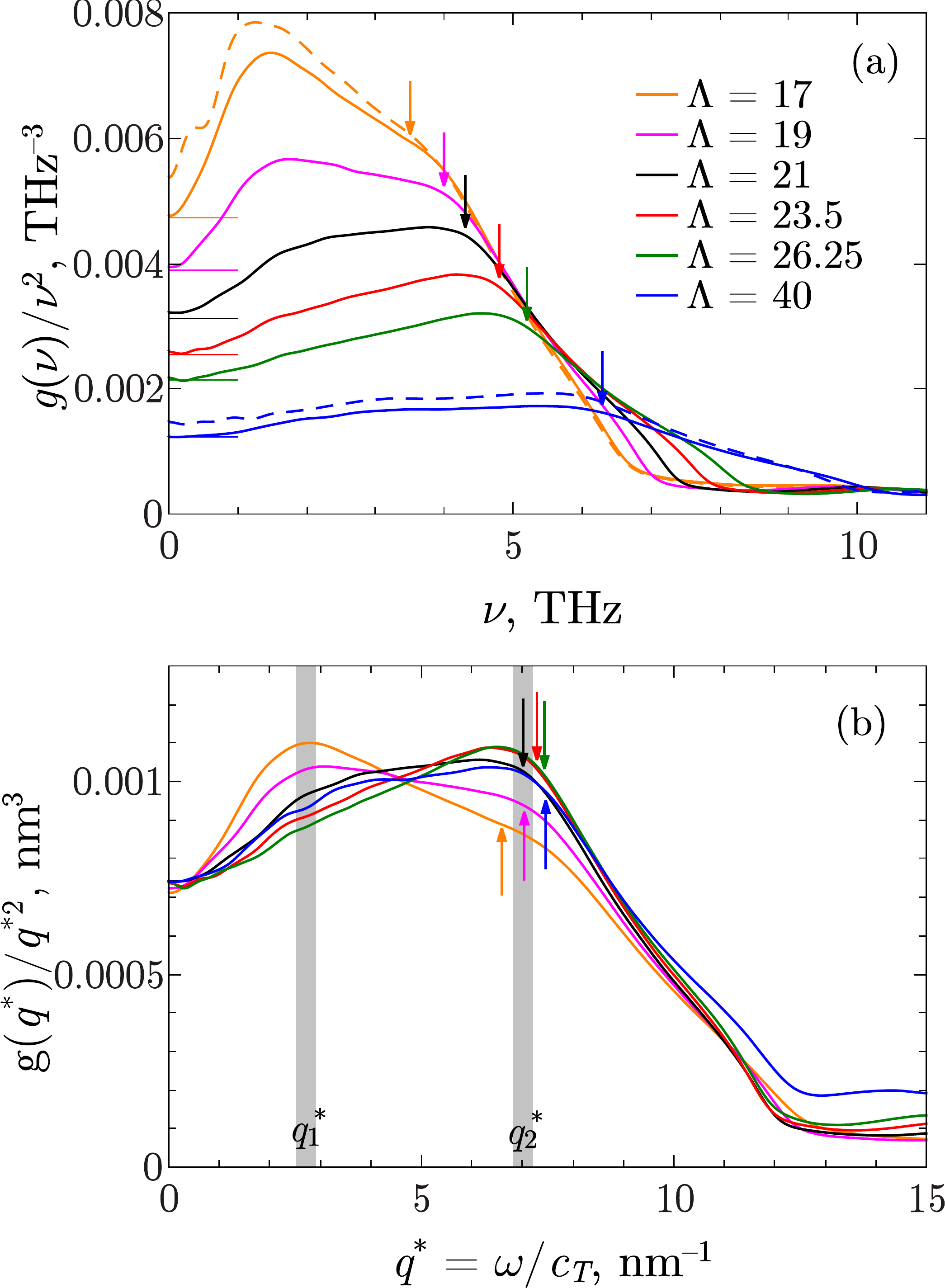}
    \caption{(a) The VDOS divided by $\nu^2$, that shows the Boson Peak for different $\Lambda$. Full lines correspond to constant density configurations as described in Table~\ref{Table1}. Dashed lines are relaxed configurations with $P\approx 0$ GPa. Horizontal thin lines show the Debye predictions calculated from the static shear and bulk modules. (b) Boson Peak as a function of the reduced wave vector $q^* = \omega/c_T$. Vertical gray bands mark the position of $q_1^*$ and $q_2^*$. Arrows show the position of the transverse Ioffe-Regel criteria (see Sec.~\ref{sec:sf}).}
\label{f.BP}
\end{figure}

The dynamical matrix $M$ has ${\cal N} = 3N$ eigenvalues that are squares of the corresponding eigenfrequencies $\omega_j$. The normalized vibrational density of states (VDOS) as a function of $\omega=2\pi\nu$ reads
\begin{equation}
    g(\omega) = \frac{1}{\cal N}\sum_{j=1}^{\cal N} \delta(\omega-\omega_j).  \label{eq:prDOS}
\end{equation}
The full set of eigenvalues for a small system (with $N < 10^4$) can be obtained by standard numerical routines. We used the FEAST Eigenvalue Solver~\cite{feast2009} for $N=8 000$ to get the full set of eigenfrequencies together with the eigenvectors of the dynamical matrix.  However, this direct method requires too much time as well as random access memory for large enough systems ($N > 10^4$). For these purposes it is necessary to use more powerful methods for the larger systems studied. In the Appendix~\ref{sec:KPM} we discuss the Kernel Polynomial Method (KPM) and velocity auto-correlation method.

The numerical results for VDOS obtained using KPM are presented in Fig.~\ref{f.DOS}. They show the usual shape of VDOS obtained for amorphous silicon~\cite{aSi-VDOS-Exp}, with a first peak related to acoustic transverse modes and a second well defined peak at high frequencies that is reminiscent of optic modes in the crystal (Fig.~\ref{f.DOS}(a)). The rescaling of the frequencies by the transverse sound velocity (Fig.~\ref{f.DOS}(b)) allows drawing the density of states as a function of a reduced wave vector $q^*\equiv\omega/c_T$. In this case, the low $q$-part of the spectra superimpose whatever the value of $\Lambda$, confirming the dominant transverse acoustic character of the low-frequency vibrations, and suggesting the existence of a characteristic length at a wave vector $q^*\approx 10$~nm$^{-1}$ independent of $\Lambda$, above which the rescaled VDOS split. The transverse sound velocities $c_T$ shown in the inset of  Fig.~\ref{f.DOS} have a $\Lambda^{1/2}$ dependence at constant density, showing that the shear modulus is proportional to the three-body contribution to the total energy of the system. This effect is not shown in the longitudinal sound velocities $c_L$ because bending rigidity is not dominant for the propagation of compressive waves, contrary to shear waves. The parameter $\Lambda$ thus allows tuning the transverse wave velocities independently of the longitudinal one.

The Boson peak is visible after dividing the VDOS $g(\nu)$ by $\nu^2$ (the Debye prediction) as shown in Fig.~\ref{f.BP}(a). The shape of the Boson peak (Fig.~\ref{f.BP}(a)) shows clearly a dependence on the bonds directionality quantified by the parameter $\Lambda$. The Boson peak appears to be magnified when the three-body interactions are low, and it decreases when the three-body interactions get more and more important as compared to the central interatomic forces. For $\Lambda=21$ corresponding to a-Si, the initial very low frequency peak is no more marked, but the Boson peak is still visible with an excess of low frequency vibrations as compared to the Debye prediction. As the value of $\Lambda$ increases, the position of the peak is shifted to higher frequencies. This effect is clearly dominated by $\Lambda$. We have checked that pressure differences between the samples induce only a small change in the Boson peak (dashed lines in Fig.~\ref{f.BP}(a)) as compared to the role of $\Lambda$. In order to quantify the observed shift to higher frequencies, we again rescale the frequencies by the transverse sound velocity, as suggested in~\cite{b.tanguy20062}. The resulting reduced density of states is shown in Fig.~\ref{f.BP}(b). The position of the Boson peak as a function of the reduced wave vector $q^*$ appears now independent of $\Lambda$, suggesting a universal process dominated by transverse waves, that will be discussed later. Note however, that the fine structure of the peak depends on the bonds directionality $\Lambda$: at a very low frequency, a peak is visible for low values of $\Lambda$, located at $q_1^*\approx 2.7$~nm$^{-1}$ (corresponding to a wavelength $\xi_1^*\approx 23$ \AA). This very low frequency peak disappears progressively and a secondary peak appears at $q_2^*\approx 7.0$~nm$^{-1}$ ($\xi_2^*\approx 9$ \AA) when $\Lambda > 21$. The significance of these peaks will be discussed later.

\begin{figure}[t]
    \centerline{\includegraphics[scale=0.38]{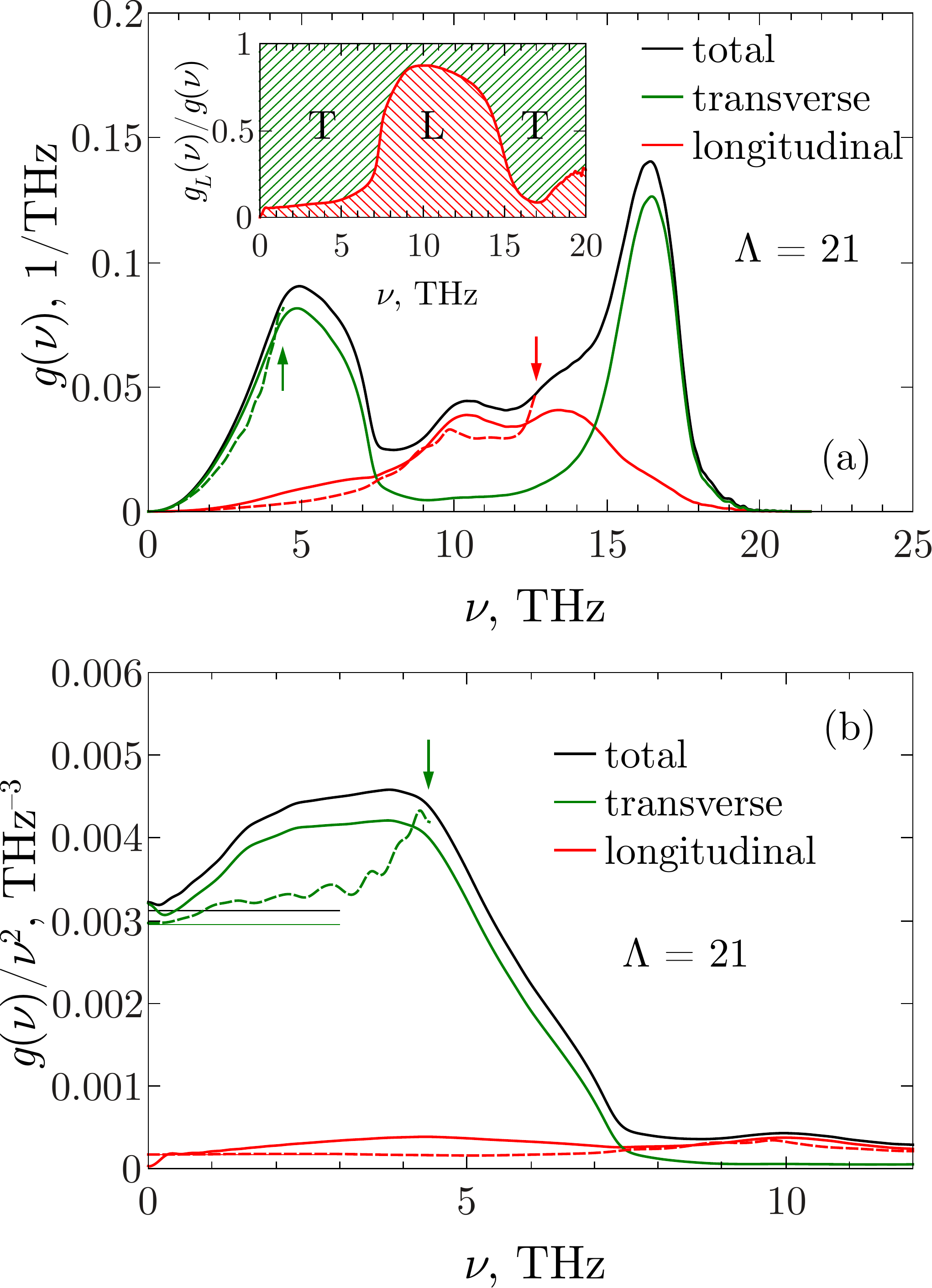}}
    \caption{(a) The decomposition of the total vibrational density of states to longitudinal and transverse components for $\Lambda=21$. The inset shows the relative number of the longitudinal modes $g_L(\omega)/g(\omega)$ (red line). The relative number of the transverse modes $g_T(\omega)/g(\omega) = 1 - g_L(\omega)/g(\omega)$ is shown by green hatching between red line and the value 1. The vertical arrows show the transverse and longitudinal Ioffe-Regel frequencies. (b) The boson peak for $\Lambda=21$ and its longitudinal and transverse components. Thin horizontal lines show the Debye prediction calculated from the static shear and bulk moduli. Dashed lines are estimations from Sec.~\ref{sec:sf} of the phononic contribution, below the Ioffe-Regel limit. The vertical arrow shows the transverse Ioffe-Regel frequency.}
    \label{f.DOSLT}
\end{figure}

Above we have shown that the low-frequency part of the VDOS has presumably a dominant transverse character. It would be very interesting to find a regular way to separate the VDOS into the longitudinal and the transverse components for the whole vibrational spectrum. In particular it gives us a possibility to show more clearly that the Boson peak has a transverse nature~\cite{schober2004}. In Appendix~\ref{sec:LT} we describe a generalized decomposition method without the notion of the wave vector, which is an ill-defined quantity in strongly disordered systems. This method is based on the volume variations of the Voronoi cells during the atomic motion. The atomic displacement of each atom can be decomposed into two components, one of them preserving the volume of the Voronoi cells. The displacements preserving the volume of each Voronoi cell are identified as transverse displacements, and the other as longitudinal displacements. The separate contribution of longitudinal and transverse displacements to the total VDOS is shown in Fig.~\ref{f.DOSLT}. In the low frequency region (below 7 THz for $\Lambda=21$), the transverse modes dominate the VDOS, thus confirming the transverse character of the vibrations in the region of the Boson peak for amorphous silicon-like samples (Fig.~\ref{f.DOSLT}(b)).
The predominance of the longitudinal modes between 7 THz and 15 THz in amorphous silicon corresponds to the gap between the upper frequency of TA modes (7.5 THz) and the lower frequency of TO modes (13.9 THz) in crystalline Si~\cite{Tubino-1972}. This frequency region in crystalline Si is totally occupied by LA and LO modes (without a gap). In amorphous Si in the same frequency region the vibrations have a small transverse component (15--20\%), in agreement with the results obtained in~\cite{b.marinov}. Certainly, there are no optical phonons with well defined wavevector in amorphous silicon due to relatively strong disorder. However, the short-range order of vibrational modes in the amorphous phase can be similar to that in the crystalline phase. At the same time, our definition of the longitudinal and transverse vibrations is local because it is based on the notion of Voronoi cells. They depend on the relative motion of neighboring atoms only.

In order to complete this description, in the next section we will calculate participation ratio and the correlation function to describe the geometrical structure of the eigenmodes that are obtained as eigenvectors of the dynamical matrix.

\section{Participation ratio and spatial correlation}
\label{sec:modes}

\begin{figure}
    \includegraphics[scale=0.4]{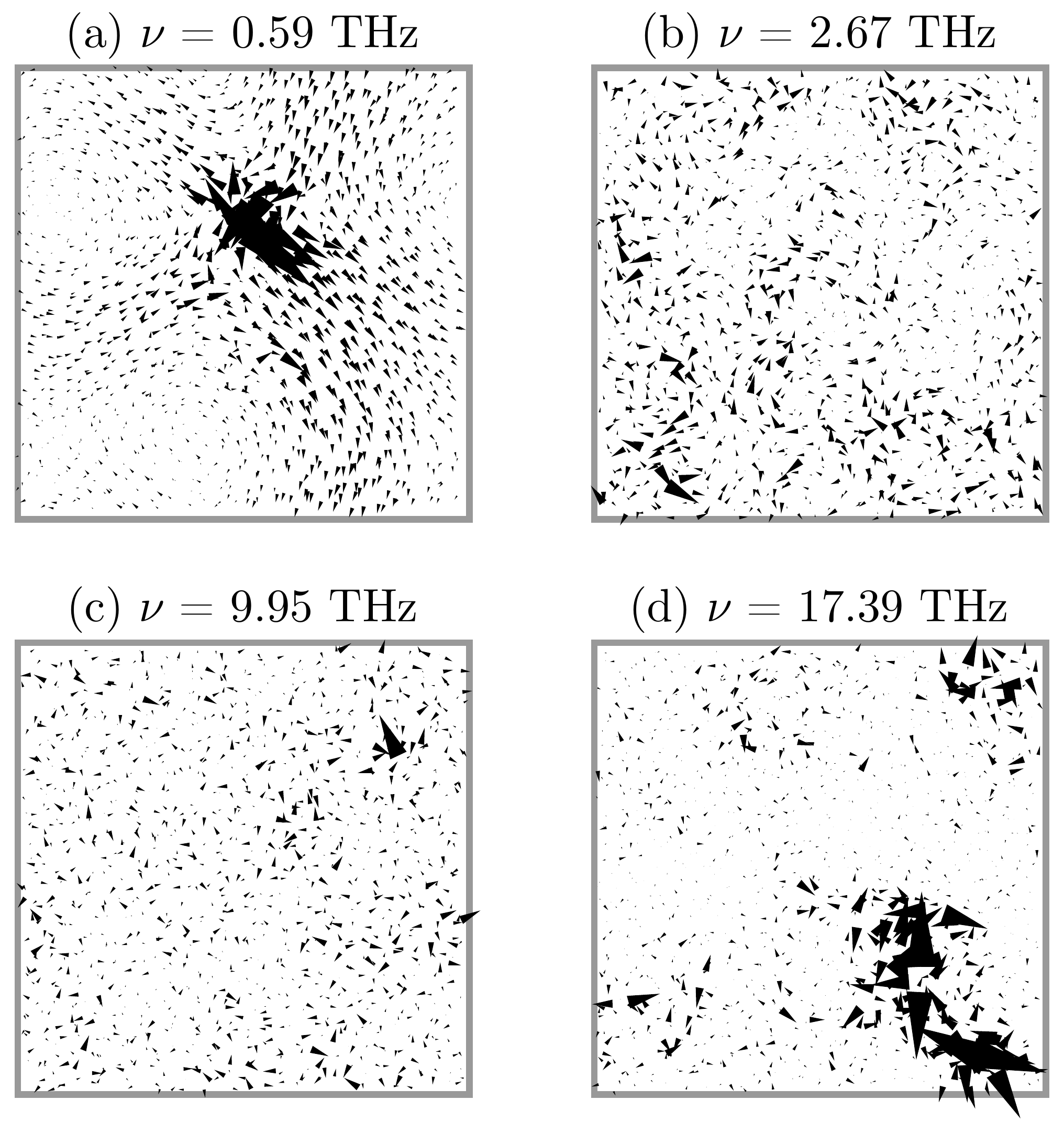}
    \caption{Vibration modes corresponding to different frequency range, for $\Lambda=21$. Arrows are proportional to the displacements of the particles ($\times100$). The 2D representation corresponds to a cut along the $x$-$y$ plane ($\delta z=5$ \AA) that contains the particle supporting the largest displacement.}
    \label{f.Visu}
\end{figure}

The exact diagonalization of the dynamical matrix is performed using FEAST Eigenvalue Solver~\cite{feast2009} on a system made with $N=8 000$ atoms. A series of ${\cal N} = 3N$ eigenmodes ${\bf u}_i(\omega_j)$ is then obtained with the corresponding eigenvalues $\omega_j^2$. These eigenmodes are the normal modes of the amorphous material. They are not phonons, because they cannot be described as simple plane waves with a well defined wave vector $\bf q$. Examples of such eigenmodes are shown in figure~\ref{f.Visu}. The low frequency eigenmodes are a superposition of plane waves with softer regions supporting highly strained isolated vibrations (Fig.~\ref{f.Visu}(a)). The modes supporting additional isolated vibrations are precursors of local plastic rearrangements when looking at the anharmonic mechanical response~\cite{b.tanguy2010}. We identify them as {\it soft modes} because they occur only in the low frequency part of the spectrum (as will be proved later) with soft spots due to very low local elastic stiffness. Other authors called the low frequency modes {\it quasi-localized modes}~\cite{b.schober, schober2004} in order to distinguish them from plane waves. At higher frequencies, the shape of the eigenmodes becomes more complex.

\begin{figure}
    \includegraphics[scale=0.4]{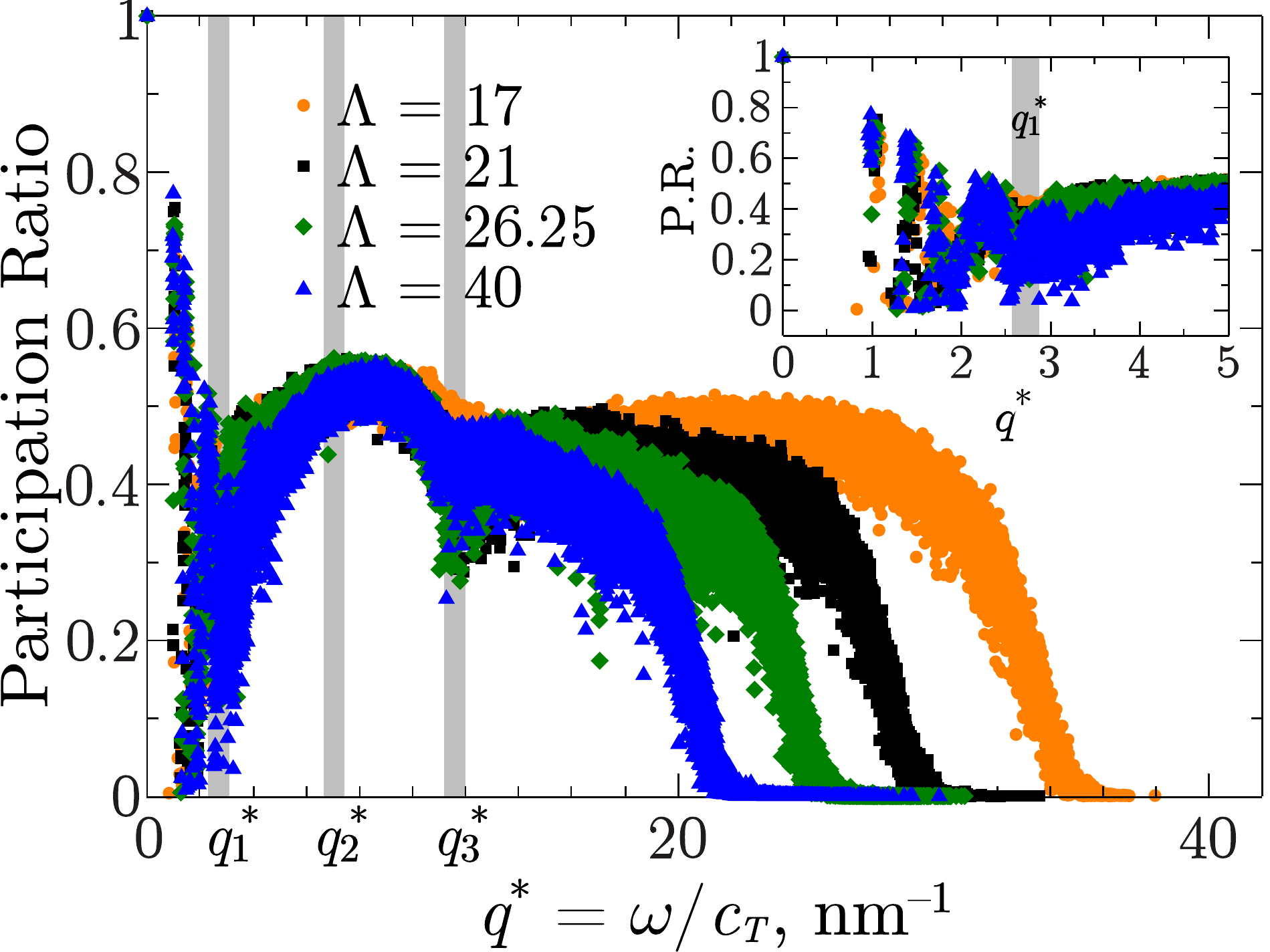}
    \caption{Participation ratio as a function of the reduced wave vector $q^* = \omega/c_T$ for different values of the parameter $\Lambda$. Vertical gray bands show three characteristics reduced wave vectors $q_1^*$, $q_2^*$, and $q_3^*$. Insets show zoom on the low-frequency range.}
    \label{f.TPart}
\end{figure}

The amount of particles moving together in the vibrational eigenmodes is usually quantified by the participation ratio defined for each eigenmode $j$ as
\begin{equation}
    P(\omega_j)\equiv\frac{1}{N}\frac{\left(\sum_i u^2_i(\omega_j)\right)^2}{\sum_i u^4_i(\omega_j)}.
    \label{eq.TP}
\end{equation}
For an isolated particle $P=1/N$, and for translational motion $P=1$. The participation ratio is shown in Fig.~\ref{f.TPart} as a function of the reduced wave vector $q^*$. Similarly to the VDOS, the low frequency part of $P$ superimposes for all $\Lambda$ when plotted as a function of the reduced wave vector $q^*$, suggesting the existence of a common geometrical origin involving mainly transverse vibrations. It can be schematized as follows: first an initial decay due to the wavelengths decrease of acoustic modes, together with very low values characteristic of soft modes. Then an increase up to a value close to $P^*=0.5$ (an example of such mode is shown in Fig.~\ref{f.Visu}(b)). The value of $P^*$ is close to 0.6 for uncorrelated Gaussian random noise~\cite{Beltukov2011}. After a secondary minimum (mode shown in Fig.~\ref{f.Visu}(c)) the participation ratio decreases to zero at the mobility edge~\cite{b.allen} that follows the position of the high frequency peak in the VDOS (see Fig.~\ref{f.DOS}). Typical mode in this frequency range is shown in Fig.~\ref{f.Visu}(d).  Quite remarkably, the position of the first minimum in the participation ratio corresponds for all $\Lambda$ to the first maximum in the rescaled VDOS divided by $\nu^2$ located at  $q_1^*$, and the position of the common maximum $P^*=0.5$ is located at $q_2^*$ corresponding to the second peak in the low frequency rescaled VDOS divided by $\nu^2$. The departure from the plane waves participation ratio in this range, means that in all our systems, the Boson peak is located in a frequency range were plane waves are no more the dominant contribution to the eigenmodes. This frequency range is limited by two characteristic distances $\xi_1^*$ and $\xi_2^*$, that are independent on $\Lambda$. There is also a third characteristic reduced wave vector $q_3^*=11.7$~nm$^{-1}$ which corresponds to a secondary local minimum of the participation ratio for all values of $\Lambda$. It coincide with the sharp change of the nature of vibrations from almost transverse to almost longitudinal ones (Fig.~\ref{f.DOSLT}).

\begin{figure}
    \includegraphics[scale=0.4]{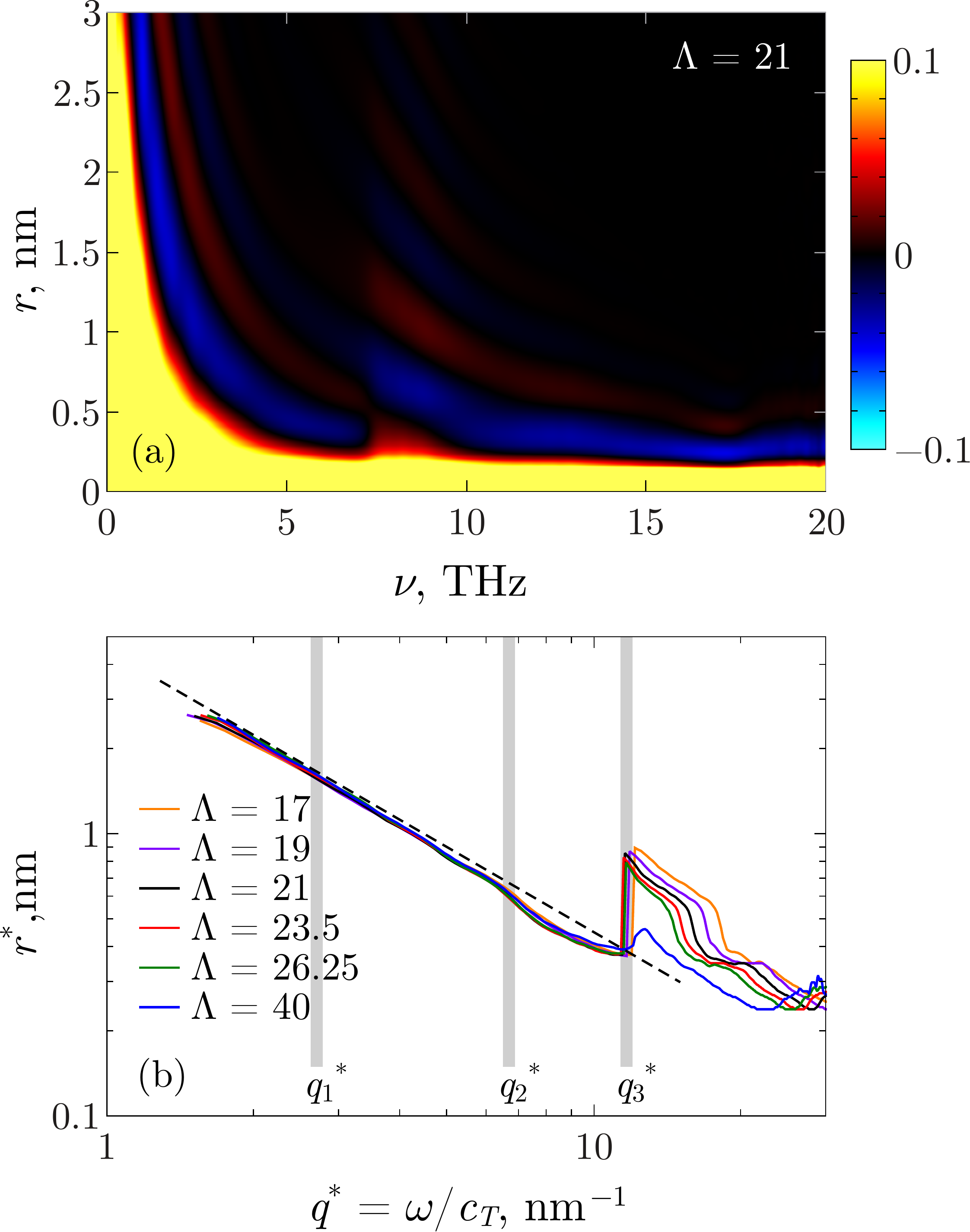}
    \caption{(a) Spatial correlation function of the atomic displacements $C_n(r,\omega)$ as a function of the frequency $\nu$ for $\Lambda=21$. Color scales indicates the amplitude of the correlation function. All amplitudes above 0.1 is shown as 0.1. (b) Position in the first minimum of the correlation function as a function of the reduced wave vector $q^*$ for the different values of $\Lambda$. The dashed line is $r^*=0.449/q^*$, which corresponds to the first minimum of Eq.~(\ref{eq:corrph}) for transverse modes. Vertical gray bands mark the positions of $q_1^*$, $q_2^*$, and $q_3^*$.}
    \label{f.Cn}
\end{figure}

In order to detail the shape of the eigenmodes and compare with these characteristic lengths, we have computed their spatial correlation function $C({\bf r}, \omega)$. It is defined as in Ref.~\cite{b.tanguy2002} as
\begin{equation}
    C({\bf r}, \omega) = \frac{1}{\cal N}\sum_{j=1}^{\cal N}\langle {\bf u}({\bf r}+{\bf r}',\omega_j)\cdot{\bf u}({\bf r}',\omega_j)\rangle_{{\bf r}'}\delta(\omega-\omega_j),  \label{eq:corr}
\end{equation}
where ${\bf u}({\bf r},\omega_j)$ is coarse-grained displacement field of $j$th eigenmode
\begin{equation}
    u({\bf r},\omega_j) = \sum_{i=1}^N {\cal W}({\bf r}-{\bf R}_i)u_i(\omega_j)
\label{eq:CG}
\end{equation}
with ${\cal W}$ a coarse-graining function normalized by $\int {\cal W}^2({\bf r})d{\bf r} = 1$. We used Gaussian coarse-graining function of width $w_{CG}=0.5$ \AA. This length is less than the typical distance between atoms, so we can neglect the overlapping of different grains. In this case, normalization of the eigenmodes implies
\begin{equation}
    \langle {\bf u}({\bf r},\omega_j)\cdot{\bf u}({\bf r},\omega_j)\rangle_{\bf r}=1,
\end{equation}
which results in the property $C(0, \omega)=g(\omega)$. For convenience we used the normalized correlation function
\begin{equation}
    C_n({\bf r}, \omega) = \frac{C({\bf r}, \omega)}{g(\omega)}.
\end{equation}

In order to calculate $C({\bf r},\omega)$ we used the KPM (Appendix~\ref{sec:evecKPM}). The amplitude of $C_n(r,\omega)$ averaged over different directions of $r$ for all the frequencies is shown in Fig.~\ref{f.Cn}(a) for $\Lambda=21$. Starting from $C_n(0,\omega)=1$, it shows oscillations between positive and negative values characterizing a spatial flipping of the displacement field. For a three-dimensional plane wave with a given polarization (L or T) and the wavevector $q$ the normalized correlation function has a form
\begin{equation}
    C_n(r,\omega) = \frac{\sin(qr)}{qr}.  \label{eq:corrph}
\end{equation}
The low-frequency behavior of the correlation function (Fig.~\ref{f.Cn}) is indeed dominated by the wavelength of the plane wave: when plotted as a function of the reduced wave vector $q^*=\omega/c_T$, it shows the characteristic behavior of transverse plane waves. Indeed, Fig.~\ref{f.Cn}(b) shows the position of the first minimum $r^*$ of $C_n$ as a function of the reduced wave vector. In the low frequency regime, it decays like  $r^*=0.449/q^*$ in exact correspondence with the wavelength of the transverse plane wave. This means that $C_n$ is dominated by the collective dynamics of plane waves even in the presence of soft modes. However, the values of $r^*=\xi_1^*$ at $q_1^*$ and $r^*=\xi_2^*$ at $q_2^*$, confirm the signature of a characteristic wavelength in the vibration modes. The origin of these lengths is not obvious. It was obvious from $P(\omega_j)$ that the eigenmodes are not simple plane waves at $q_1^*$, but have a very small participation ratio indicating the presence of isolated centers of enhanced vibrations. These centers are sufficiently few to not affect the long range spatial correlations due to transverse plane waves in $C_n$, but can affect the vibrational density of states through small frequency shifts. This description supports the fact that transverse plane waves and isolated centers of enhanced vibrations still coexist at $q^*_1$. $\xi^*_1$ could be the distance between the isolated centers, that would correspond as well to the wavelength at $q^*_1$. At $q^*_2$ a clear-cut change of behaviour appears in $C_n$ for all the systems studied indicating departure from transverse plane waves. This effect will be discussed again later. Finally a common change appears at a larger reduced wave vector $q^*_3=11.7$~nm$^{-1}$ in $C_n$, which was introduced in Fig.~\ref{f.TPart}. It corresponds to the transition from transverse modes to longitudinal ones with bigger correlation radius.

The analysis of the geometrical structure of the eigenmodes would not be complete without analysing the scaling dependence of the participation ratio. This kind of analysis is especially important when looking at the signature of localization in the Anderson interpretation~\cite{vanTiggelen}. For that, we computed the generalized participation ratio $P_k$ of order $k$ defined as
\begin{equation}
    P_k(\omega_j,w_{CG})\equiv\frac{1}{N}\frac{\left(\int u_{CG}^2({\bf r}, \omega_j)d{\bf r}\right)^k}{\int u_{CG}^{2k}({\bf r}, \omega_j)d{\bf r}}
    \label{eq.TPk}
\end{equation}
where ${\bf u}_{CG}({\bf r}, \omega_j)$ is the displacement field of mode $j$ coarse-grained at different scales $w_{CG}$ as in Eq.~\ref{eq:CG}.

\begin{figure}
    \includegraphics[scale=0.3]{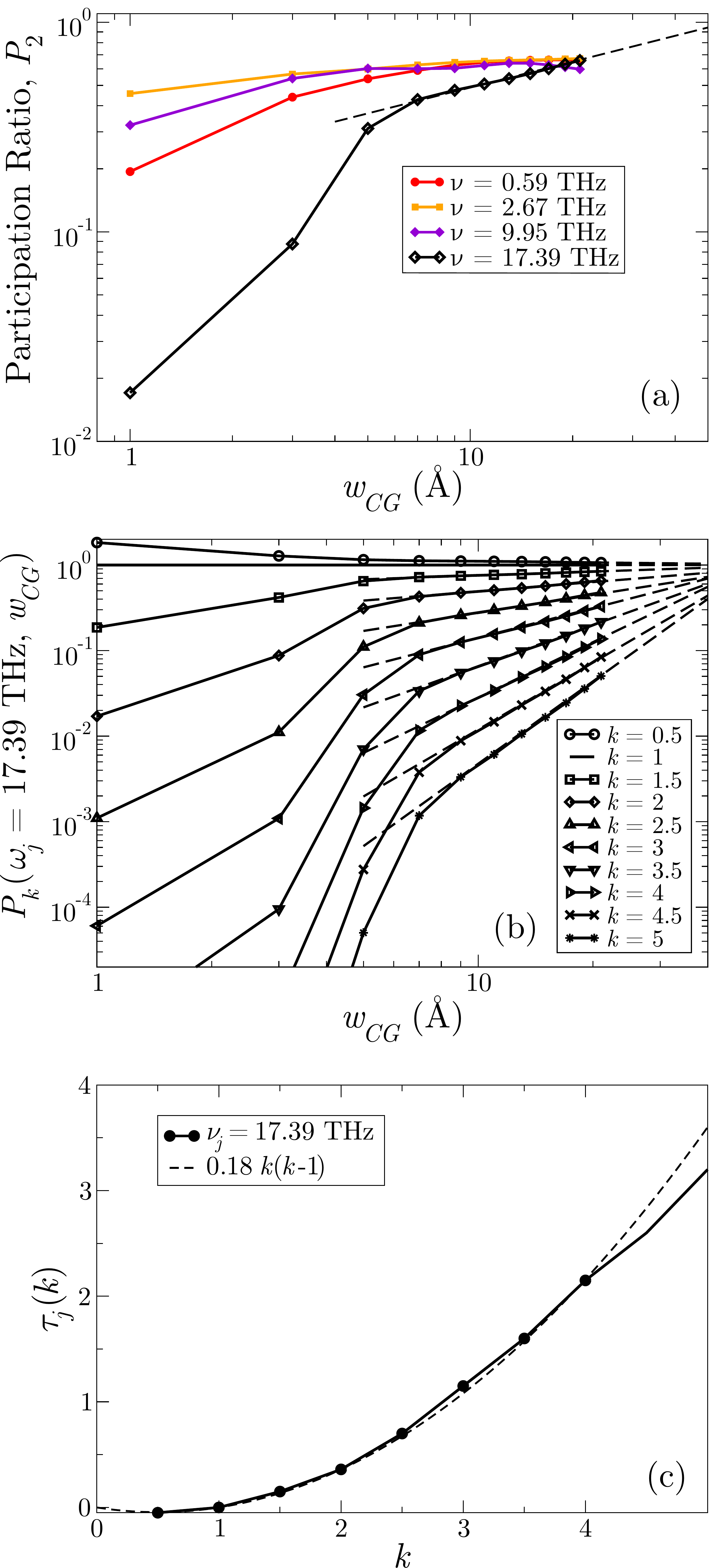}
    \caption{(a) Coarse-grained participation ratio $P_2(\omega_j,w_{CG})$ as a function of the coarse-graining scale $w_{CG}$ for different frequencies $\omega_j$ including those of  the vibration modes shown in Fig.~\ref{f.Visu}, with $\Lambda=21$. (b) Generalized participation ratio $P_k(\omega_j,w_{CG})$ for $\nu_j= 17.39$ THz and $k$ in $ \{0.5,1,...,5\}$. Dashed lines are power-law fits. (c) Exponent $\tau_j(k)$ for $\nu_j=17.39$ THz in the mobility edge, compared to the quadratic fit proposed in the context of localization theory~\cite{Localization}.}
    \label{f.TPart-CG}
\end{figure}

The dependence of $P_2$ as a function of $w_{CG}$ shown in Fig.~\ref{f.TPart-CG}(a) reveals a size invariance (power-law behaviour) only for the largest frequency modes studied, that is above  the mobility edge (black curve). In opposite, the participation ratio $P_2$ of very low frequency soft mode (red curve in Fig.~\ref{f.TPart-CG}(a)) goes down for $w_{CG}< w_{CG}^*\approx 9$\,\AA, but saturates at large $w_{CG}$, confirming the coexistence of enhanced local vibrations with large scale plane waves in this frequency range. Note that the corresponding size of the soft spot $w_{CG}^*\approx\xi_2^*$, and the upper saturation occurs at $w_{CG}^{**}=2\pi/q^*\approx\xi_1^*$. In the mobility edge (Fig.~\ref{f.TPart-CG}(b)) scale invariance manifests through the power-low behaviour $$P_k(\omega_j,w_{CG})\propto w_{CG}^{\tau_j(k)}$$ for large $w_{CG}$. We measured the exponent $\tau_j(k)$ as a function of $k$ (dashed lines in Fig.~\ref{f.TPart-CG}(b)) for a mode $n$ in the mobility edge. It is shown in Fig.~\ref{f.TPart-CG}(c) and it confirms the localized and multi-fractal~\cite{Page2009,Localization} hallmark of this mode, with
\begin{equation}
    \tau_j(k) \approx \gamma k(k-1)\quad\text{and}\quad\gamma=0.2.
\end{equation}

We have shown in this section, that the eigenmodes have characteristic features depending on the corresponding frequency range. In the low frequency part of the spectrum, eigenmodes share common features independent on the bending rigidity of the modes: for example, the Boson peak is bounded by two characteristic lengthscales $\xi^*_1$ and $\xi^*_2$ with a very low participation ratio in the first case, and a local maximum in the participation ratio in the second case. These two length-scales have a signature in the spatial correlation analysis of the modes. It is shown as well, that in the very low frequency range, transverse plane waves coexist with local enhanced vibrations, while the plane wave character of the vibrations is questioned in the higher limit of the Boson peak. In the mobility edge, at higher frequencies, eigenmodes have a multifractal shape characteristic of a localized behavior. We will now study the dynamical structure factor, in order to relate these observations to the study of density-density correlation functions, as can be tracked in neutron diffraction experiments for example~\cite{giordano_2010,giordano_2011}. The analysis of the dynamical structure factor allows also to discuss the Ioffe-Regel criterion for waves scattering.

\section{Dynamical Structure Factor\label{sec:sf}}

\begin{figure}[t]
    \includegraphics[scale=0.4]{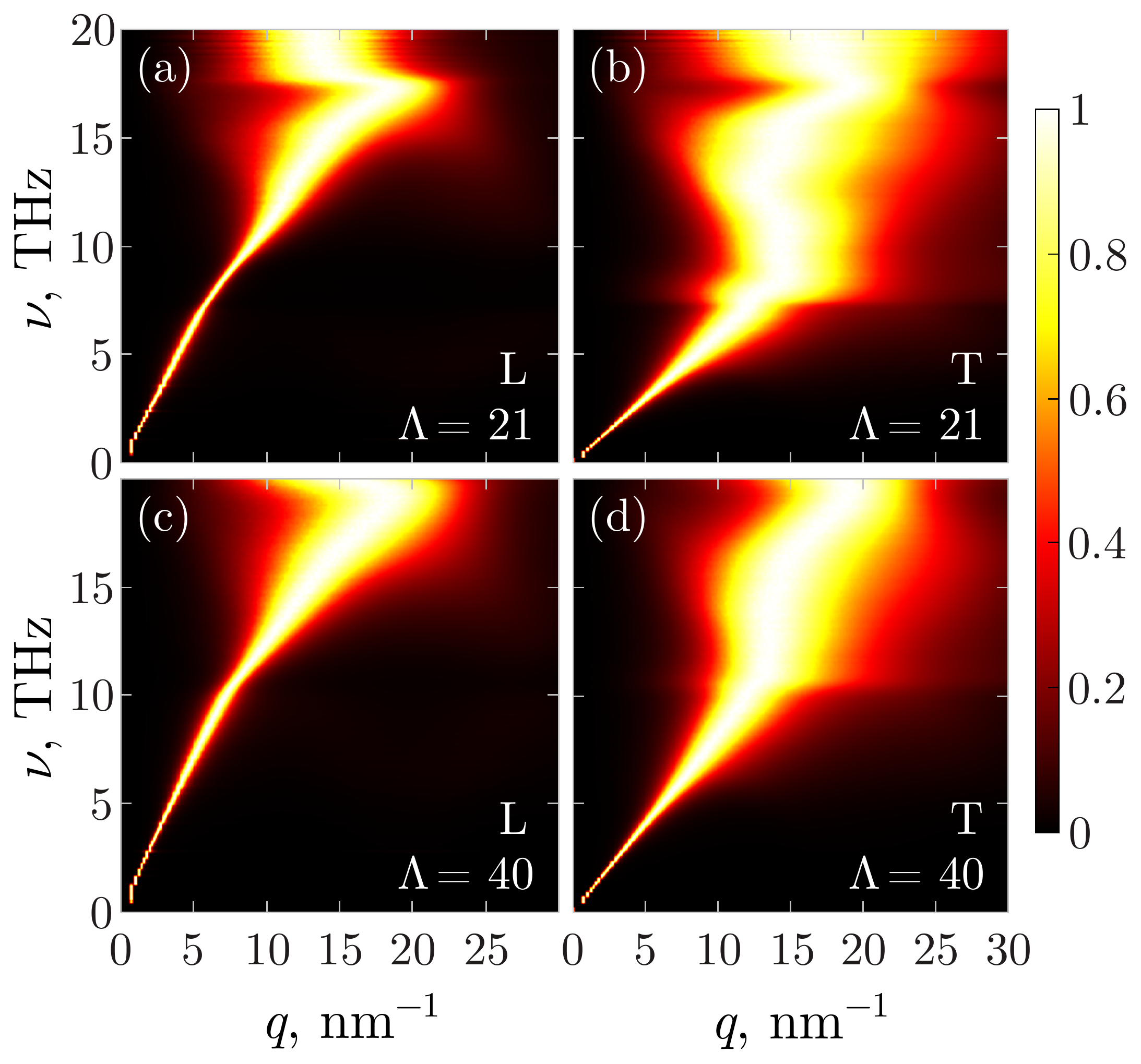}
    \caption{The longitudinal (L) and transverse (T) dynamical structure factors as a function of the wavenumber $q$ and the frequency $\nu$ for the parameter $\Lambda = 21$ and $\Lambda = 40$.}
    \label{f.SF}
\end{figure}

In this section we analyze the dynamical structure factor in order to determine the dispersion law and the mean free path for longitudinal and transverse phonons. We prove that it is an accurate method in the frequency range below the Ioffe-Regel criterion, where mean free path is still bigger than the half wavelength, and the notion of phonon dispersion is well defined.

The dynamical structure factor is the self-correlation function of the mass currents~\cite{shintani} in the system at thermal equilibrium with some temperature $T$. In the small temperature limit $T\to 0$ the structure factor can be calculated by normal mode analysis
\begin{equation}
    S_\eta({\bf q}, \omega) = \frac{k_BT}{m}\frac{q^2}{\omega^2}F_\eta({\bf q}, \omega), \label{eq:S}\\
\end{equation}
where $\eta$ denotes longitudinal (L) or transverse (T) component. In the above equation $F_\eta({\bf q}, \omega)$ is the longitudinal or transverse component of the Fourier transform of the eigenmodes
\begin{align}
    F_L({\bf q}, \omega) &= \sum_{j=1}^{\cal N}\biggl|\sum_{i=1}^N\hat{\bf q}\cdot{\bf u}_i(\omega_j)e^{i{\bf q}{\bf R}_i}\biggr|^2\delta(\omega-\omega_j),   \label{eq:Fl} \\
    F_T({\bf q}, \omega) &= \sum_{j=1}^{\cal N}\biggl|\sum_{i=1}^N\hat{\bf q}\times{\bf u}_i(\omega_j)e^{i{\bf q}{\bf R}_i}\biggr|^2\delta(\omega-\omega_j). \label{eq:Ft}
\end{align}
Here $\hat{\bf q} = {\bf q}/|{\bf q}|$ is a unit vector along ${\bf q}$ and ${\bf u}_i(\omega_j)$ is the displacement of the $i$th atom for $j$th eigenmode.

In order to calculate $F_\eta({\bf q}, \omega)$ we also used the KPM (Appendix~\ref{sec:evecKPM}). Fig.~\ref{f.SF} shows the structure of eigenmodes $F_\eta(q,\omega)$ in the reciprocal space averaged over all possible directions of $\bf q$. For a better visual effect we divide $F_\eta(q,\omega)$ by the magnitude of its maximum for each fixed value of $\omega$. All color maps in Fig.~\ref{f.SF} have two evident regions: low-frequency region with thin phonon branch, and high-frequency region without a certain relationship between the wavenumber $q$ and the frequency $\omega$.

\begin{figure}
    \includegraphics[scale=0.4]{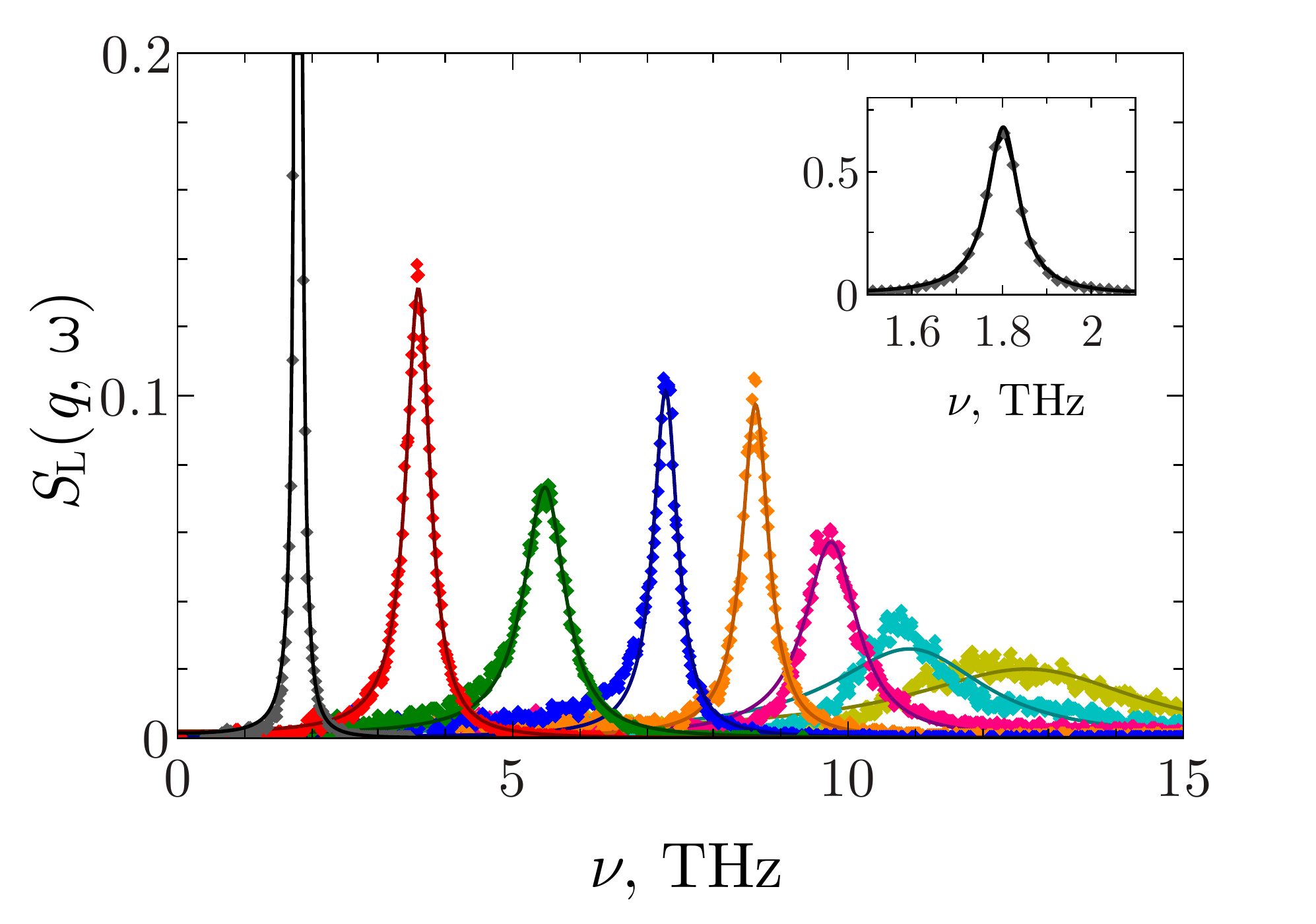}
    \caption{Fits of the dynamical structure factor $S_L(q,\omega)$ to Eq.~(\ref{eq:fit}) for $\Lambda=21$ and various values of the wavenumber $q$ (from left to right: 1.44, 2.89, 4.33, 5.77, 7.21, 8.66, 10.10 and 11.54~nm${}^{-1}$). The inset shows a full curve for $q=1.44$~nm${}^{-1}$.}
    \label{f.fit}
\end{figure}

In order to extract information about phonons in the low-frequency region we fit the structure factor $S_\eta(q,\omega)$ using the DHO model (Fig.~\ref{f.fit})
\begin{equation}
    S_\eta(q,\omega) = \frac{A}{(\omega^2-\omega_\eta^2(q))^2+\omega^2\Gamma^2}, \quad \eta=L,T.   \label{eq:fit}
\end{equation}
We extract phonon dispersion $\omega_\eta(q)$, the phonon inverse life time $\Gamma(q)$ and a coefficient $A$ from this fit.

\begin{figure}
    \includegraphics[scale=0.4]{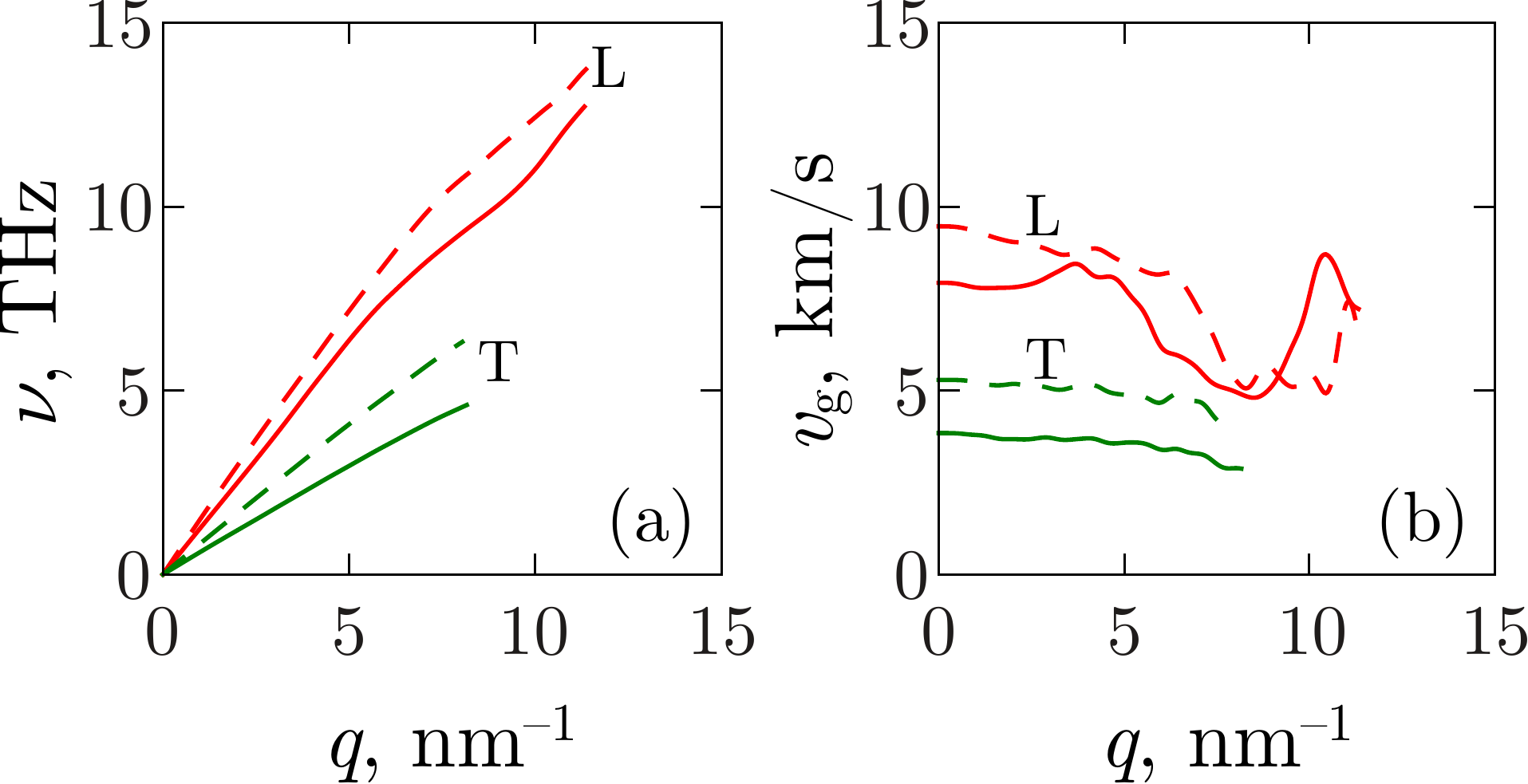}
    \caption{(a) Phonon dispersion curve obtained from the fitting (\ref{eq:fit}) for the parameter $\Lambda=21$ (full line) and $\Lambda=40$ (dashed line). (b) Group velocity for the parameter $\Lambda=21$ (full line) and $\Lambda=40$ (dashed line). All curves are shown up to the corresponding Ioffe-Regel frequency. Longitudinal and transverse phonons are denoted by L and T respectively.}
    \label{f.disp}
\end{figure}

\begin{figure}
    \includegraphics[scale=0.4]{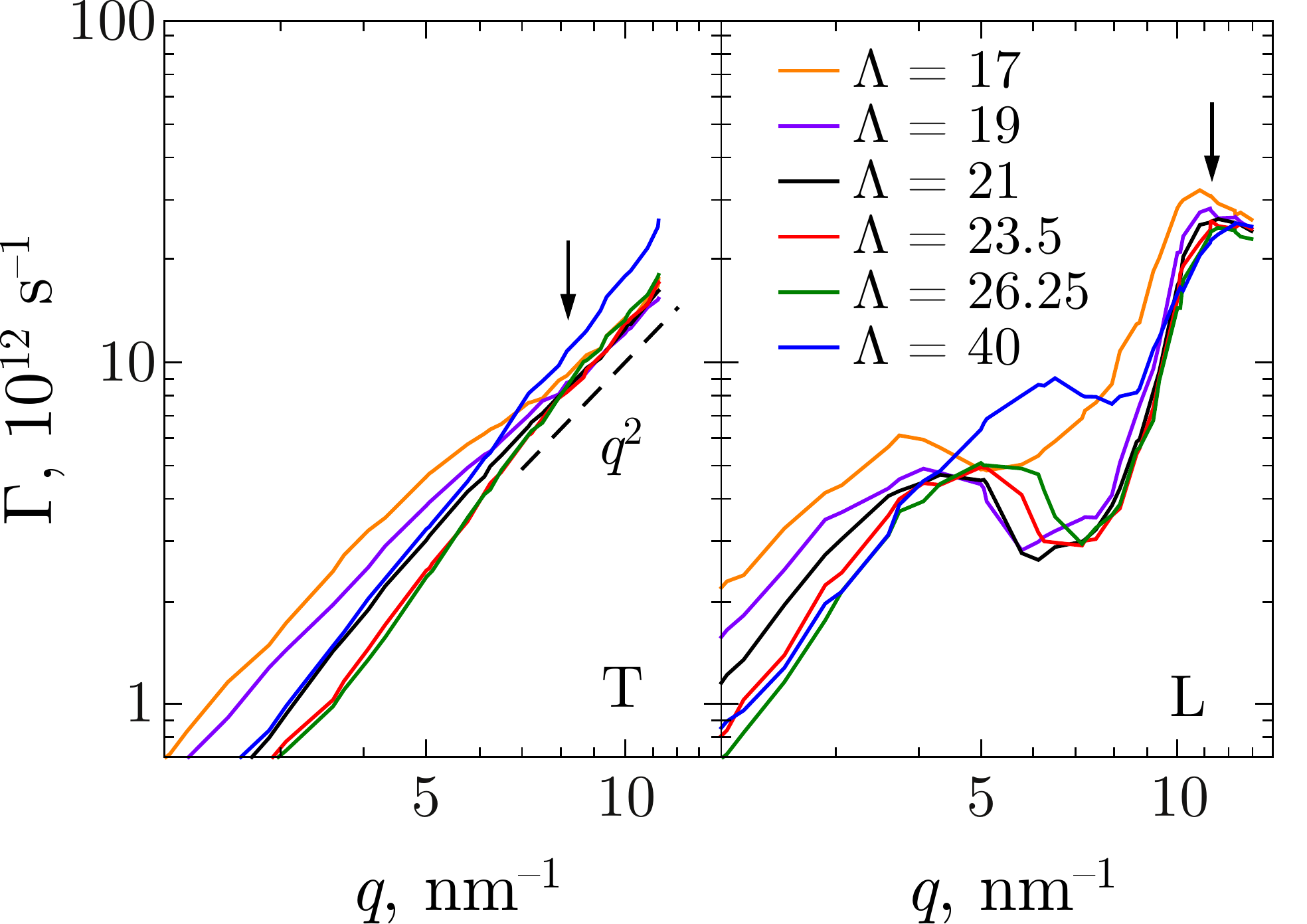}
    \caption{Width $\Gamma$ obtained from the DHO fit of the structure factor, as a function of the wave vector $q$ obtained from the dispersion relation Fig.~\ref{f.disp} for the different values of $\Lambda$. Left: transverse modes; right: longitudinal modes. Arrows mark the Ioffe-Regel criterion for all values of $\Lambda$. Transverse modes show the $\Gamma\propto q^2$ law near the Ioffe-Regel criterion.}
    \label{f.Gq}
\end{figure}

\begin{figure}
    \includegraphics[scale=0.4]{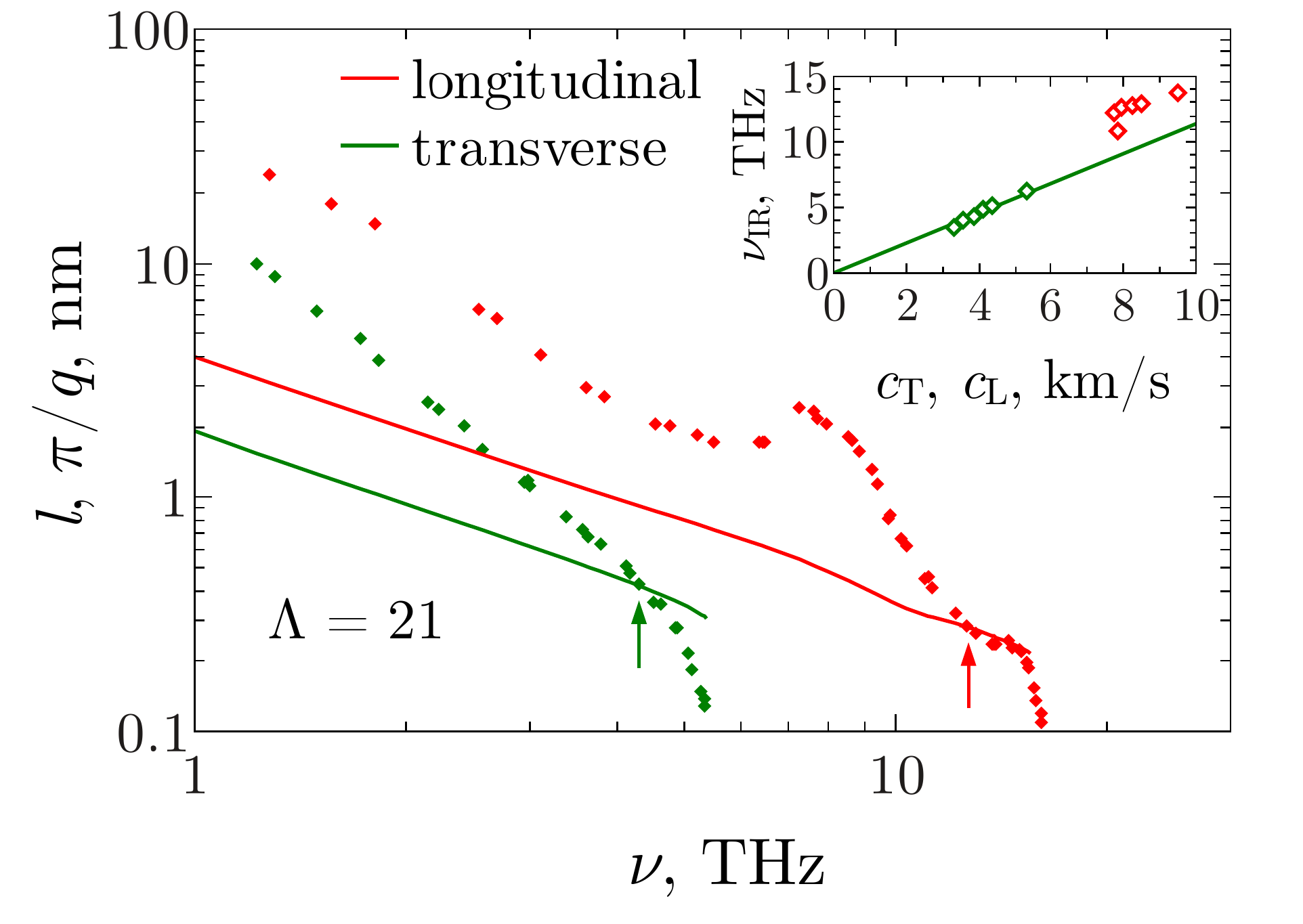}
    \caption{The mean-free path $l$ as a function of the frequency (points), compared to the half-wavelength $\pi/q$ given by DHO fit (solid lines). The crossing points determines the Ioffe-Regel criterion (shown by arrows). Inset: the Ioffe-Regel frequencies for longitudinal and transverse phonons for different values of the parameter $\Lambda$. The solid line shows the trend $\nu_{\rm IR}^T = c_T/\xi_2$.}
    \label{f.IR}
\end{figure}

The numerical results obtained by this method for phonon dispersion $\omega_\eta(q)$ as well as the group velocity $v_g^\eta = \partial \omega_\eta/\partial q$ are presented in Fig.~\ref{f.disp}. With a known value of $\Gamma$ and $v_g$, the phonon mean free path $l(\omega)$ can now be calculated as $l(\omega) = v_g \Gamma$. The phonons are well defined excitations if their mean free path $l(\omega)$ exceeds the phonon half wavelength $\pi/q$ (Ioffe-Regel criterion for phonons~\cite{Beltukov2013}). Fig.~\ref{f.Gq} shows the value of $\Gamma$ for all the samples with different bending rigidities $\Lambda$ and Fig.~\ref{f.IR} shows the position of the Ioffe-Regel frequency for longitudinal and transverse phonons for the parameter $\Lambda = 21$. From the similar figures for other values of the parameter $\Lambda$ we find the remaining Ioffe-Regel frequencies (Table \ref{t.IR}).

\begin{table}[b]
    \caption{Transverse and longitudinal Ioffe-Regel criteria for different values of the parameter $\Lambda$.}
    \begin{tabular}{*{6}{@{\hspace{0.2cm}}c@{\hspace{0.2cm}}}}
        \hline\hline
        $\Lambda$ & $\nu_{\rm IR}^T$, THz & $\nu_{\rm IR}^L$, THz & $q^{T}_{\rm IR}$, nm$^{-1}$ & $q^{L}_{\rm IR}$, nm$^{-1}$  \\
        \hline
        17    & 3.3 & 12.1 & 6.2 & 9.7\\
        19    & 4.1 & 12.4 & 7.2 & 10.1\\
        21    & 4.5 & 12.7 & 7.3 & 10.0\\
        23.5  & 5.1 & 12.8 & 7.8 & 9.8\\
        26.25 & 5.7 & 13.0 & 8.2 & 9.6\\
        40    & 7.0 & 13.8 & 8.3 & 9.1\\
        \hline\hline
    \end{tabular}
    \label{t.IR}
\end{table}

Different comments are raised by these measurements. First, the sound wave velocities (Fig.~\ref{f.disp}(b)) are well defined below the Ioffe-Regel criterion. It is not constant, but it varies with $q$. Transverse sound velocities show a small decay with $q$, starting at the low frequency limit of the Boson peak, as already measured in experiments~\cite{giordano_2011}. Longitudinal sound velocities decay faster with a sudden increase at $q\approx 10$~nm${}^{-1}$ ($\nu=12.5$ THz for $\Lambda=21$), after transverse waves became strongly scattered in the sample.

The density of states of longitudinal and transverse phonons can be obtained from the dispersion law
\begin{align}
    g_L(\omega) &= \frac{L^3}{6N\pi^2}\frac{q_L(\omega)^2}{v_g^L(\omega)}, \\
    g_T(\omega) &= \frac{L^3}{3N\pi^2}\frac{q_T(\omega)^2}{v_g^T(\omega)}.
\end{align}
and compared to the more general decomposition that was already discussed in Sec.~\ref{sec:DOS}. In the low-frequency limit the ratio between them is
\begin{equation}
    g_L(\omega)/g_T(\omega) = c_T^3/2c_L^3\ll1.    
\end{equation}
For $\Lambda=21$ this ratio is equal to 0.057, which coincides well with the low frequency part of this ratio shown in the inset in the Fig.~\ref{f.DOSLT}(a). The very low frequency modes are naturally mainly transverse due to their lower sound velocity. The two estimations of longitudinal and transversal contribution to VDOS are compared in Fig.~\ref{f.DOSLT}. The estimation from the dispersion law is close to the general estimation, but slightly lower. The main difference is on the Boson peak Fig.~\ref{f.DOSLT}(b): the first low frequency peak (attributed to soft modes in Sec.~\ref{sec:DOS}) is indeed completely absent in the dispersion law estimation of the VDOS that stays close to the Debye one in this frequency range. It means that this first peak results from a departure to the phonon-like behaviour.

The inverse life time $\Gamma$ (Fig.~\ref{f.Gq}) is different for transverse and for longitudinal waves. The inverse life time of transverse waves varies approximately $\propto q^2$ as discussed extensively in the litterature~\cite{Ruffle2006}, with a collapse for all $\Lambda$ values at $q^*_2$. Longitudinal inverse life time is more sparse. It shows a sudden increase at $q\approx 10$~nm$^{-1}$, that is after transverse waves are strongly scattered in the system and do not interfere anymore with longitudinal waves.
The measurement of $\Gamma$ allows computing a mean-free path $l$ (Fig.~\ref{f.IR}) but only in the region were the sound velocity $v_g$ is well defined. The computed longitudinal mean-free path is always larger than the transverse mean-free path. The measured values of $l$ can overcome the size of the system, because it is computed from an estimation of the inverse life time $\Gamma$ that does not result from a propagating process~\cite{Damart2015} but only from a general fit of a geometrical function $S(q,\omega)$. The Ioffe-Regel cross-over occurs at a well defined wave vector $q^T_{\rm IR}=2\pi\nu_{\rm IR}^T/c_T\approx 7.5\pm 1.0$~nm$^{-1}$ for transverse waves, and $q^L_{\rm IR}=2\pi\nu_{\rm IR}^L/c_L\approx 9.7\pm 0.4$~nm$^{-1}$ for longitudinal waves (see Table~\ref{t.IR}), slightly larger than the upper limit $q^*_2$ of the Boson peak for the transverse one. This relation is only slightly sensitive to $\Lambda$, suggesting a universal mechanism for strong scattering in amorphous materials, independent on the specific interatomic interactions.
We will now compare this estimation of the Ioffe-Regel criterion to the description of quasi monochromatic wave packet propagation.

\section{Diffusivity\label{sec:diff}}

\begin{figure}[t]
    \includegraphics[scale=0.4]{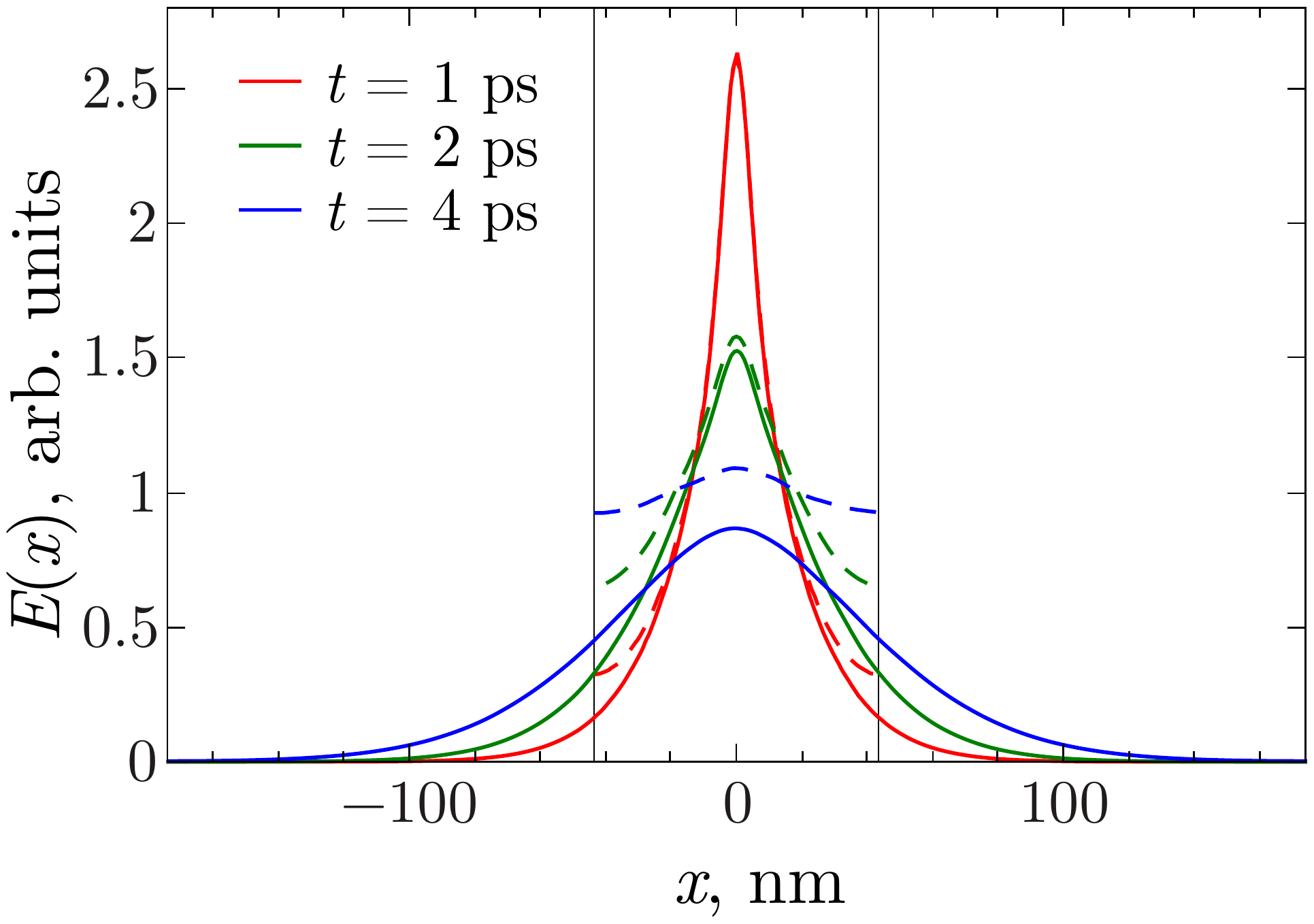}
    \caption{The spatial distribution of the vibrational energy for different time $t$ after the maximum of the exciting pulse for $\nu = 8$ THz. Solid lines were calculated for the repeatedly extended sample with $4\times2\times2$ periodic blocks. Dashed lines show the spreading of the energy in one periodic block only. Vertical thin black lines shows boundaries of one periodic block. Horizontal plot range corresponds to the width of the extended sample.}
    \label{f.Edistr}
\end{figure}

In this section we consider the diffusion of the vibrational energy. For this purpose we excite a quasi monochromatic wave packet in the middle thin layer of the sample around $x=0$ in a small time interval around $t=0$. To excite vibrations in the sample we use the excitation force
\begin{equation}
    f_{i\alpha}^{\rm ext}(t) = \sin(\omega t+\varphi_{i\alpha})\exp\left(-\frac{x_i^2}{2w^2}-\frac{t^2}{2T^2}\right)   \label{eq:ext}
\end{equation}
where the phase $\varphi_{i\alpha}$ is random for each atom $i$ and each Cartesian projection $\alpha$. The width of the excited layer is determined by the value of $w=3$~nm and the duration of the excitation is given by $T = 1$ ps. The latter determines the frequency resolution $\Delta\nu=1$ THz. We start our calculations at time $t_0=-5T$ when the external force is still negligible. In order to have a sufficiently large system size, the central sample with periodic boundary conditions and size $L=87$\,\AA is duplicated into 4 images along $x$ direction and 2 along $y$ and $z$ directions. As a result in volume size we obtain a 16 times bigger sample. This allows a determination of large mean free paths for phonons and diffusion of energy on longer distances. Indeed, the energy diffusion front reaches the boundaries of the original sample at $t\approx 1$ ps when the excitation force is still active (see dashed lines in Fig.~\ref{f.Edistr}). The diffusivity of the vibrational energy in this extended sample is the same as in one big sample except a small region near the mobility edge. Localized modes with the localization length $\xi>L$ look like delocalized in the repeated sample.

\begin{figure}
    \includegraphics[scale=0.7]{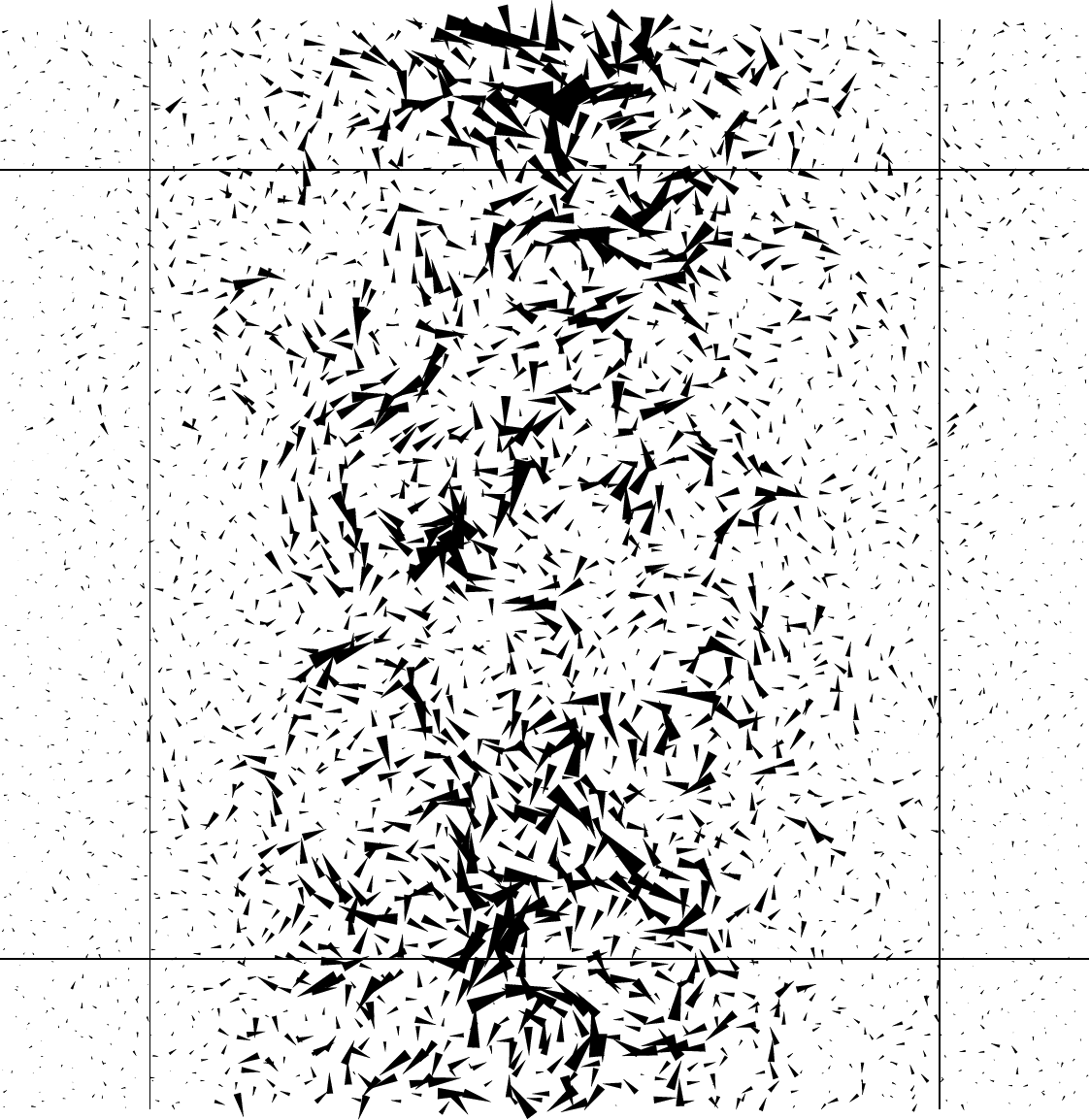}
    \caption{Snapshot of rotons with wave packet for $\Lambda=21$, $\nu=4$ THz and $t = 2$ ps.}
    \label{f.SnapWP}
\end{figure}

\begin{figure}[t]
    \includegraphics[scale=0.4]{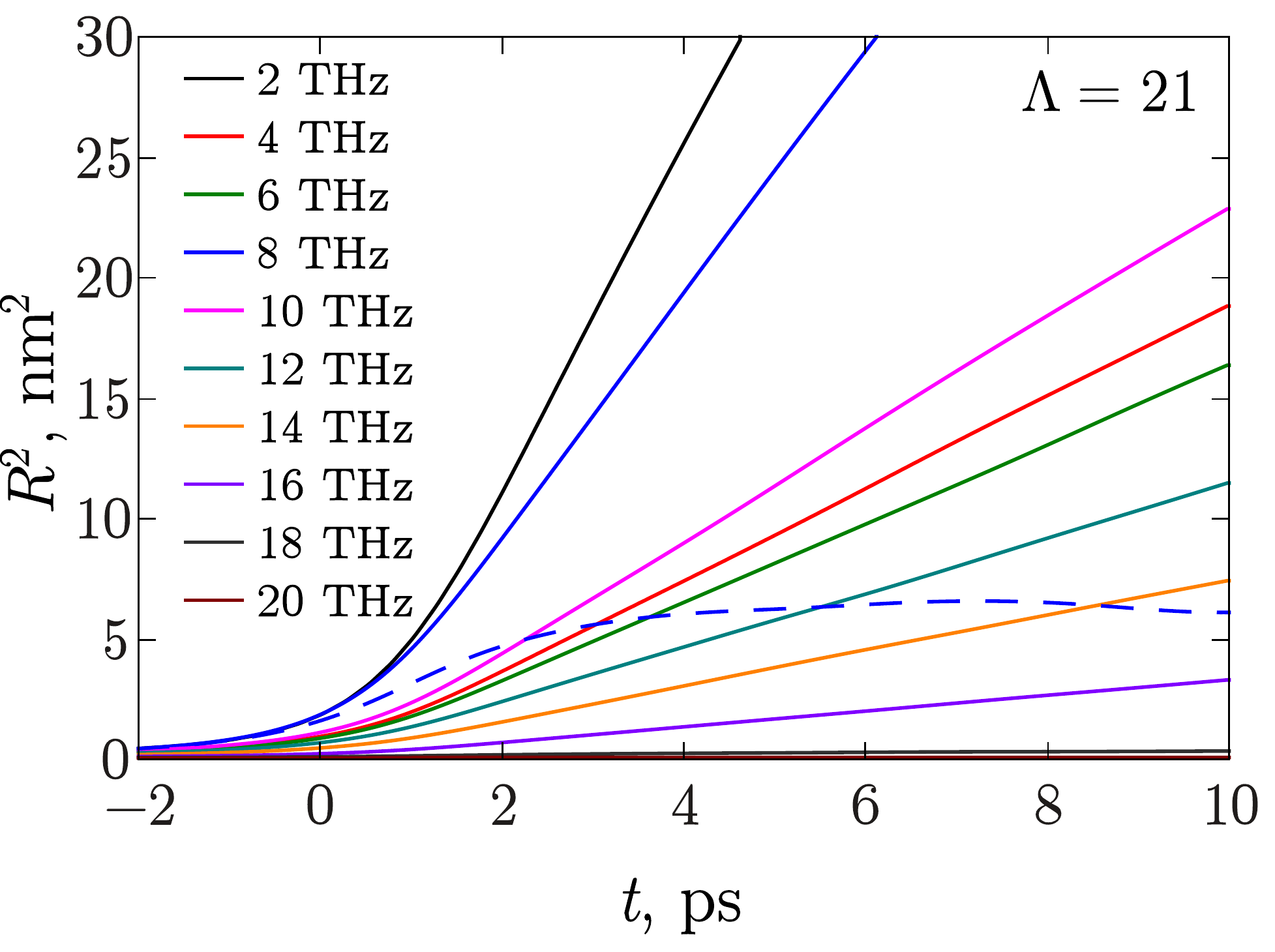}
    \caption{Spreading of the vibrational energy in space $R^2(t)$ for different frequencies for $\Lambda=21$. Solid lines were calculated for the repeatedly extended sample with $4\times2\times2$ periodic blocks. Dashed line show the spreading of the energy in one periodic block only (for $\nu = 8$ THz). }
    \label{f.R2}
\end{figure}

\begin{figure}
    \includegraphics[scale=0.4]{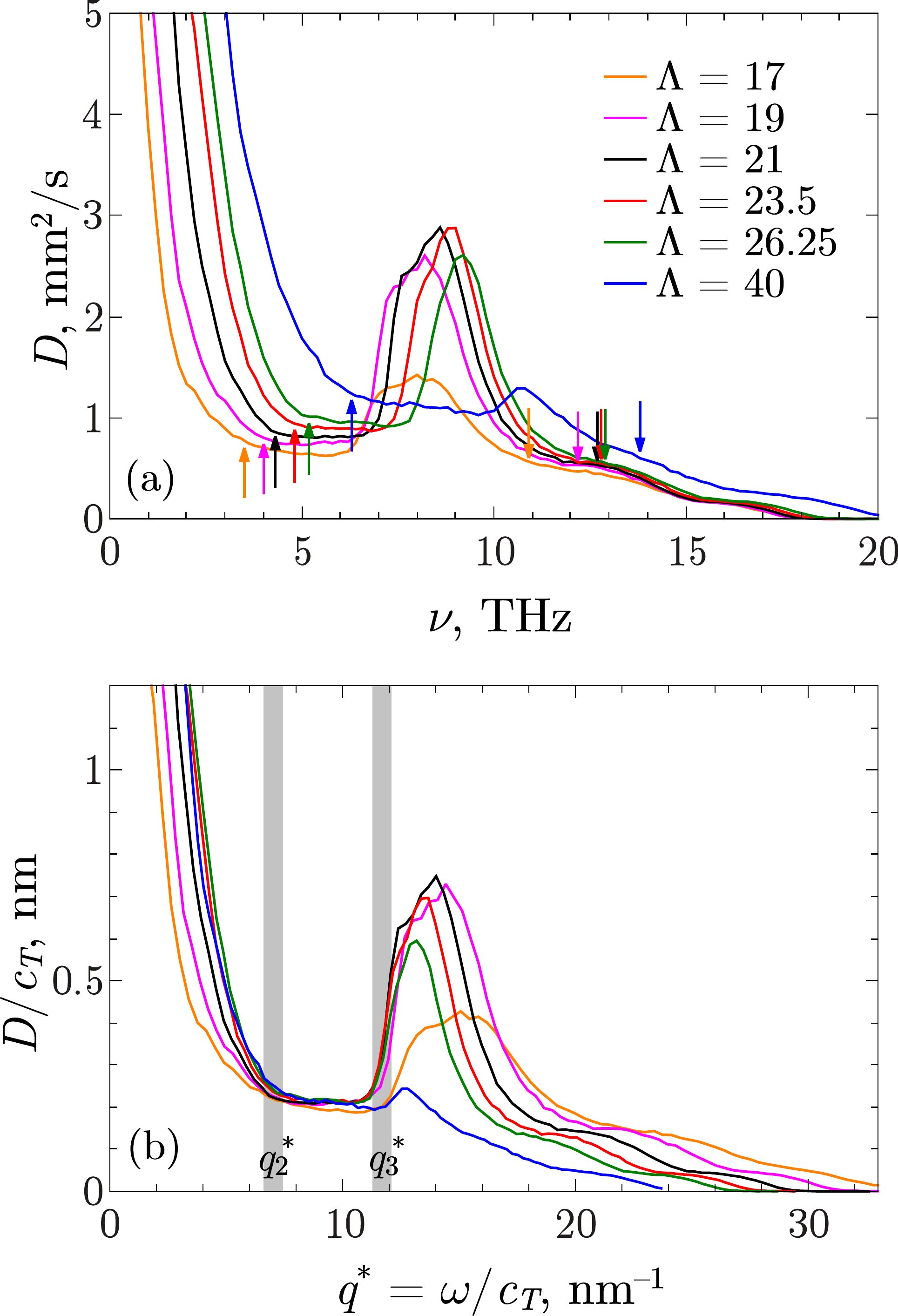}
    \caption{(a) The diffusivity as a function of frequency for different values of the parameter $\Lambda$. Upward and downward arrows show the transverse and the longitudinal Ioffe-Regel criteria respectively. (b) The rescaled diffusivity $D/c_T$ as a function of the reduced wave vector $q^* = \omega/c_T$ for the same values of~$\Lambda$. Vertical gray bands mark the positions of $q_2^*$ and $q_3^*$.}
    \label{f.Diff}
\end{figure}

\begin{figure}
    \centerline{\includegraphics[scale=0.38]{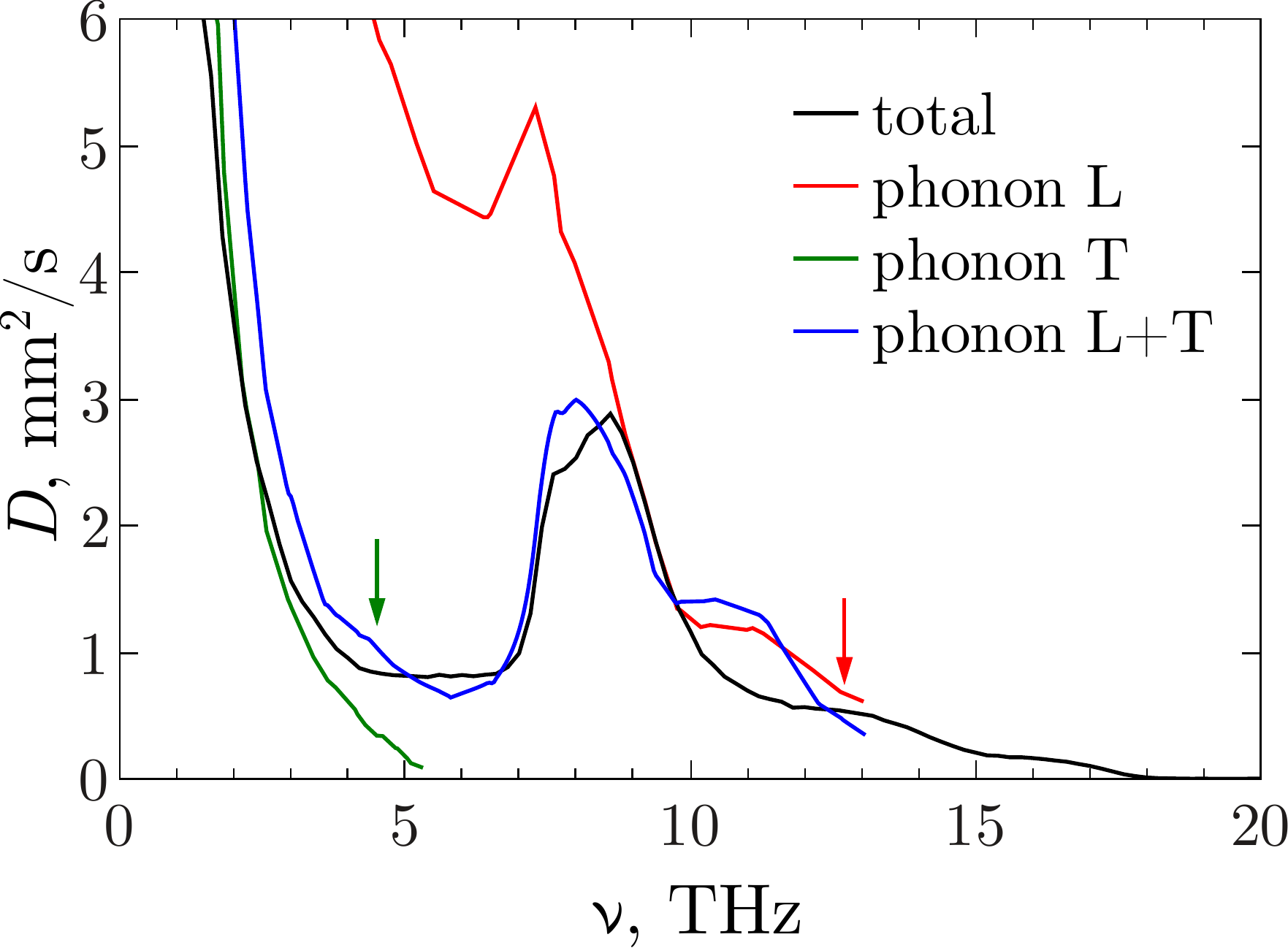}}
    \caption{Diffusivity of longitudinal phonons (red line) and transverse phonons (green line). Blue line shows the estimation of phonon contribution to the total diffusivity by Eq.~(\ref{eq:DiffPh}). Black line is the total measured diffusivity. The vertical arrows show the transverse and longitudinal Ioffe-Regel frequencies.}
    \label{f.DiffLT}
\end{figure}

After applying the external force, the vibrational energy spreads in both directions from the central layer $x=0$. The average radius squared of the energy diffusion front is defined as in Ref.~\cite{Beltukov2013} as:
\begin{equation}
    R^2(t) = \frac{1}{E_{\rm tot}}\sum_i x_i^2 E_i(t).
\end{equation}
Here, $x_i$ is the $x$ coordinate of the $i$th atom, $E_i(t)$ is the total energy of the $i$th atom, and sum is taken over all atoms in the sample. $E_{\rm tot} = \sum_iE_i(t)$ is the total vibrational energy of the system. It is independent on time, after the external force $f_{i\alpha}^{\rm ext}(t)$ became negligibly small (i.e., for $t>5T$).

The energy of the $i$th atom $E_i(t)$ is the sum of the kinetic energy and a half of the potential energy of connected bonds with $i$th atom:
\begin{equation}
    E_i(t) = \frac{v_i(t)^2}{2}-\frac{1}{2}\sum\limits_{j\alpha\beta} M_{i\alpha,j\beta}u_{i\alpha}(t)u_{j\beta}(t).
\end{equation}
Here, ${\bf v}_i(t)=\dot{\bf u}_i(t)$ is the $i$th atom velocity with the same notations as Eq.~\ref{eq:motion}.
The spatial vibrational energy distribution along the $x$ direction is shown in Fig.~\ref{f.Edistr} at different times $t$. Initial random phases $\varphi_{i\alpha}$ allow keeping the center of mass of the energy in the central layer, while the energy is progressively spread inside the sample.

To integrate the system with a given external force and zero initial conditions we used the Verlet method with a small enough time step $\delta t=0.6$ fs, and get the dependence $R^2(t)$ for different frequencies from $\nu = 2$ THz up to $\nu = 20$ THz. The results are shown in Fig.~\ref{f.R2}. We clearly see a linear temporal dependence in each curve. Their slope gives us the diffusivity by the equation for one-dimensional diffusion
\begin{equation}
    R^2(t) = 2D(\omega)t.
\end{equation}
The resulting diffusivity is shown in Fig.~\ref{f.Diff}(a) for different values of the parameter $\Lambda$. All curves have the same structure: (1) low-frequency modes with large diffusivity; (2) a flat region with relatively small diffusivity; (3) a prominent peak of the diffusivity; (4) a gradual decreasing of the diffusivity; (5) zero diffusivity for localized modes. The first two regions coincide for all values of $\Lambda$ if we plot the rescaled diffusivity $D/c_T$ as a function of the reduced wave vector $q^* = \omega/c_T$ (Fig.~\ref{f.Diff}(b)).
After an initial decay, the diffusivity saturates at a minimum value. Whatever $\Lambda$, the flat region in the diffusivity occurs precisely between $q^*_2$ and $q^*_3$, that is in the region between the upper bound of the Boson peak (close to the Ioffe-Regel criterion for transverse waves) and transition from mostly transverse modes to mostly longitudinal ones. The lower boundary of the flat region is in perfect agreement with those obtained in Ref.~\cite{Beltukov2013} for a completely different random system, where it was shown that the minimum in the diffusivity coincides with the Boson peak, at a frequency corresponding to the Ioffe-Regel criterion for transverse waves. The relation between Boson peak and Ioffe-Regel criterion was also suggested by experimental measurements~\cite{Ruffle2006}, and molecular dynamics simulations in Lennard-Jones glasses~\cite{b.tanguy20061}. It is shown here that the strong scattering gives rise to a very low diffusivity, and that it is possible to measure diffusivity in a purely harmonic model as soon as interactions are random. The shape of the instantaneous velocity field in the flat region is shown in Fig.~\ref{f.SnapWP} during the propagation of a wave packet. Rotational structures are clearly visible, and responsible for the strong dephasing close to the Ioffe-Regel crossover. The flat region in the diffusivity is followed by a peak already discussed in Ref.~\cite{b.allen}.

The peak of the diffusivity is large for almost all values of the parameter $\Lambda$ (Fig.~\ref{f.Diff}).
We can find alternatively the diffusivity of longitudinal and transverse phonons up to the Ioffe-Regel criteria using the approximate relation~\cite{Kittel-book}
\begin{equation}
    D_\eta(\omega) = \frac{1}{3}l_\eta(\omega)v_\eta(\omega), \quad \eta=L,T   \label{eq:DiffLT}
\end{equation}
where $\omega<\omega_{\rm IR}^L$ for longitudinal phonons and $\omega<\omega_{\rm IR}^T$ for transverse phonons. We can see that the diffusivity of longitudinal and transverse phonons has monotonically decreasing behavior except the negligible peak for longitudinal phonons (red and green lines in Fig.\ref{f.DiffLT}). Eq.~\ref{eq:DiffLT} cannot give the diffusivity beyond the Ioffe-Regel criteria but we expect a small diffusivity decreasing down to 0 at the mobility edge. Therefore the peak in the diffusivity cannot be explained by the diffusivity of longitudinal and transverse phonons separately. However, the total diffusivity depends on the ratio of density of states of longitudinal and transverse vibrations
\begin{equation}
    D(\omega)=\frac{g_L(\omega)}{g(\omega)}D_L(\omega)+\frac{g_T(\omega)}{g(\omega)}D_T(\omega).    \label{eq:DiffPh}
\end{equation}
The resulting phononic diffusivity is shown in Fig.~\ref{f.DiffLT}. It shows that the main peak located at $q^*\approx 13.5$~nm$^{-1}$ is due to the large density of longitudinal modes $g_L(\omega)$, enhancing the small diffusivity increase due to the absence of transverse modes in that frequency range.
The rise of the diffusivity at 7 THz in amorphous silicon thus corresponds to the sharp change of the nature of vibrations from almost transverse to almost longitudinal ones having high sound velocity.

\section{Conclusion\label{sec:Conclusion}}

\begin{figure}
    \begin{center}
    \includegraphics[scale=0.4]{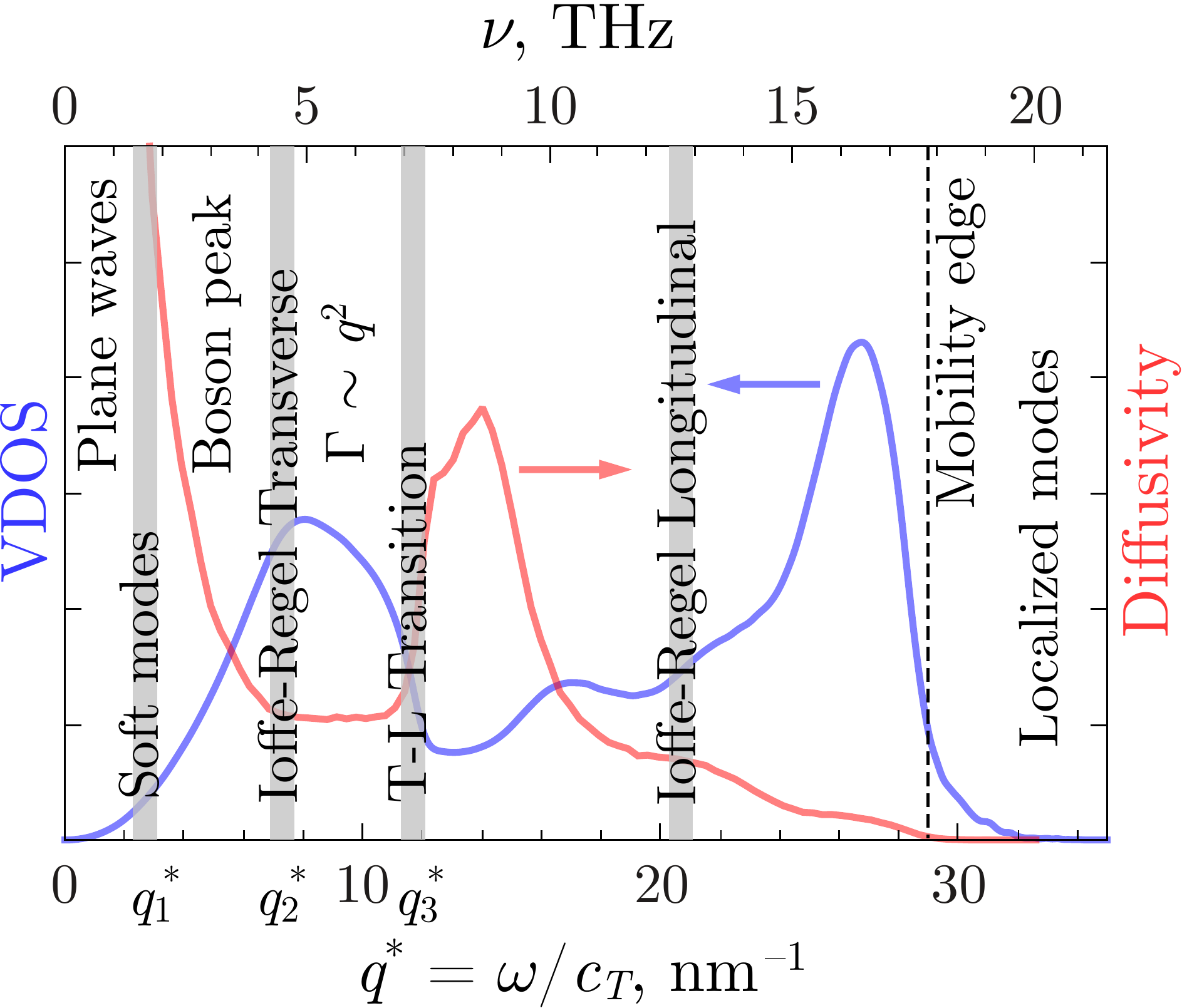}
    \caption{Schematic description of the different crossovers in the vibrational properties of harmonic amorphous solids. VDOS and diffusivity of vibrations of model amorphous silicon with $\Lambda=21$ are shown in the background.}
    \label{f.Summary}
    \end{center}
\end{figure}

We have proposed in this article a coherent picture of the vibrational properties in harmonic amorphous solids with local tetrahedral order, by combining four independent approaches: the detailed study of the normal modes (resonant vibrational modes) and of the vibrational density of states, dynamic structure factor calculation and propagation of a quasi monochromatic wave packet. The bending rigidity of local interatomic bonds was used as a control parameter to tune the sound velocity. This allowed to get a coherent picture of the vibrational response of our model systems. Different regimes were highlighted. The results are summarized in Fig.~\ref{f.Summary}.

The low frequency vibrational response is dominated by transverse modes. In this region, the Boson peak is visible and bonded by two characteristic wave vectors: the first is related to soft modes, and the second to the Ioffe-Regel limit for transverse waves. Remarkably, these two wave vectors are independent on the details of the interactions in the different systems studied here, and they define two characteristic mesoscopic  length-scales $\xi_1^*$ and $\xi_2^*$ having a signature in the spatial correlations of the normal modes. In silicon-like samples, the large difference between the transverse and longitudinal sound velocities yields a large gap between the Ioffe-Regel limit for transverse waves, and the Ioffe-Regel limit for longitudinal waves. In this gap, the vibrations sharply change the transverse character to longitudinal one near $q^*\approx q^*_3$, resulting in a deep increase of the diffusivity. As shown already by P.B. Allen at al.~\cite{b.allen}, the mobility edge and the transition to localized acoustic modes occur at higher frequencies. This transition was obtained as well in other disordered model materials, like lattice models~\cite{taraskin2003} and models of amorphous silica~\cite{taraskin2005}, thus supporting its universal feature. The modes in the Boson peak range preceding the Ioffe-Regel crossover for transverse waves have a characteristic random rotational structure yielding a dephasing of the wave front. In this frequency range, the participation ratio is at a maximum value, but the diffusivity is at a minimum value, and the diffusive inverse life time is proportional to $q^2$. Similar results have already been obtained with Brillouin scattering measurements~\cite{Ruffle2006}. However, the precise sensitivity of the inverse life times or of the vibrational density of states to the rescaled wave vector (or equivalently to the frequency) is system dependent~\cite{larkin}. Among other new results, our work shows the coherence of this picture for different systems with local tetrahedral order, thus supporting a universal geometrical process for waves scattering in amorphous systems, with three different mesoscopic characteristic wavelengths.

Perspectives of this work include the possibility to enlarge the Boson peak region (and thus the heat capacity~\cite{Kittel-book}) by increasing the first characteristic length $\xi^*_1$ related to the distance between soft spots, or decreasing the characteristic lengths $\xi^*_2$ related to the size of the strongly scattering structures, for example increasing the temperature. Another perspective is to reduce the gap between transverse and longitudinal Ioffe-Regel criteria by reducing the longitudinal sound velocity (decreasing the bulk modulus for example), in order to embed the diffusivity's raise, and thus to decrease the diffusivity in the gap (and consequently the heat conductivity~\cite{Kittel-book}). The characteristic lengthscales evidenced in this article are related to the size and distance between soft spots, and to the separation between longitudinal and transverse vibrations. However, the understanding of the structural origin of these sizes is still a challenging problem, that opens new perspectives.

\begin{acknowledgments}
We thank U. Buchenau, V. Giordano, B. Ruffl\'e and M. Wyart for very interesting discussions. This work was supported by the French Research National Agency programs MECASIL (ANR-12-BS04-0004) and Labex IMUST RAMASIL. One of the authors (Y.M.B.) thanks the Dynasty Foundation for the financial support. Two authors (D.A.P. and Y.M.B.) thanks the University Lyon~1 for hospitality.

\end{acknowledgments}

\appendix

\section{Calculation of the VDOS by KPM\label{sec:KPM}}

There are several methods for calculation of the VDOS without exact diagonalization of the dynamical matrix. Following standard statistical mechanics methods~\cite{Dove}, one can make a molecular dynamics simulation of a large system excited at a small temperature $T$. Then assuming the equipartition of the energy, we can compute VDOS as a Fourier transform of the velocity auto-correlation function~\cite{Sampoli-1998}
\begin{equation}
    g_\textsc{cvv}(\omega) = \frac{2}{\cal N}\frac{m}{kT}\int_0^{t_{\rm max}} \sum_i \overline{{\bf v}_i(t){\bf v}_i(0)} \cos \omega t\, dt.
    \label{eq:cvvdos}
\end{equation}
Here ${\bf v}_i(t)$ is the instantaneous velocity of the $i$th atom and $m$ is the atomic mass (all atomic masses are supposed to be the same). This method is much faster than a numerical diagonalization of the matrix $M$. However, it requires accurate integration of the equations of motion up to a large enough time $t_{\rm max}$ with small enough time step $\delta t \ll 1/\omega_{\rm max}$ and low temperature $k_B T<\Delta E$. Here $\omega_{\rm max}$ is the maximum frequency in the system, and $\Delta E$ the smallest energy barrier surrounding the referred equilibrium position. The resulting frequency resolution of the density of states $g_\textsc{cvv}(\omega)$ in this method is $\delta\omega\sim1/t_{\rm max}$.

The kernel polynomial method~\cite{Weisse-2006} (KPM) is an alternative way to compute the VDOS for large systems. It is a more accurate and much faster method in comparison with the previous one, as will be discussed below. It makes it possible to find the VDOS using Eq.~(\ref{eq:prDOS}) with $\delta$-function replaced by a series of polynomials. KPM was introduced in~\cite{Silver-1997} and detailed reviewed in~\cite{Weisse-2006}. Originally it was used for finding electronic DOS in disordered systems. It allows  with controlled accuracy getting directly the distribution of the eigenvalues of some large matrix $M$, not calculating numerically the eigenvalues itself. We will show how KPM can be adopted to find the VDOS, i.e. the distribution of the {\it square roots} of the eigenvalues of the dynamical matrix $M$ without its exact diagonalization. In this method we use only moments of this matrix up to sufficiently high order which is controlled by the accuracy of the calculations. Below we shortly describe the KPM for our problem.

All eigenvalues $\omega_j^2$ of the matrix $M$ are non-negative due to mechanical stability of the system and lie in some interval $[0,\omega_{\rm max}^2]$. Usually the precise value of the maximum frequency is unknown so $\omega_{\rm max}$ is an estimation of the maximum frequency which guaranties that $\omega_j<\omega_{\rm max}$ for all $\omega_j$. Let us introduce new dimensionless variable $\varepsilon = 1-2\omega^2/\omega_{\rm max}^2$ in order to rescale all eigenfrequencies squared $\omega_j^2$ to the interval $[-1,1]$ for variable $\varepsilon_j$. Thus, we can transform Eq.~(\ref{eq:prDOS}) as
\begin{equation}
    g(\omega) = \frac{4\omega}{{\cal N} \omega_{\rm max}^2}\sum_{j=1}^{\cal N} \delta(\varepsilon - \varepsilon_j)  \label{eq:prDOS2}
\end{equation}
where $\varepsilon_i = 1-2\omega_i^2/\omega_{\rm max}^2$ are eigenvalues of the matrix $\tilde{M} = I-2M/\omega_{\rm max}^2$ where $I$ is the unit matrix.

For $-1<\varepsilon<1$ and $-1<\varepsilon_j<1$ we can expand the $\delta$-function in Eq.~(\ref{eq:prDOS2}) in second kind Chebyshev polynomial series
\begin{equation}
    \delta(\varepsilon-\varepsilon_j) = \frac{2}{\pi}\sqrt{1-\varepsilon^2}\sum\limits _{k=0}^{\infty}U_k(\varepsilon)U_k(\varepsilon_j).  \label{eq:dec1}
\end{equation}
Chebyshev polynomials of the second kind are defined by recurrence relations
\begin{align}
    U_0(\varepsilon) &= 1,   \label{eq:Tk1}\\
    U_1(\varepsilon) &= 2\varepsilon,   \label{eq:Tk2}\\
    U_k(\varepsilon) &= 2\varepsilon U_{k-1}(\varepsilon)-U_{k-2}(\varepsilon).   \label{eq:Tk3}
\end{align}
They have an equivalent trigonometric definition
\begin{equation}
    U_k(\varepsilon) = \frac{\sin((k+1)\arccos \varepsilon)}{\sqrt{1-\varepsilon^2}}.   \label{eq:Tktrig}
\end{equation}
From Eqs. (\ref{eq:prDOS2}) and (\ref{eq:dec1}) the density of states can be expressed in terms of the sine Fourier transform
\begin{equation}
    g(\omega)=\frac{8\omega}{\pi\omega_{\rm max}^2}\sum_{k=0}^{\infty}\mu_k\sin ((k+1)\varphi)   \label{eq:dec2}
\end{equation}
where $\varphi$ depends on $\omega$ as $\varphi = 2\arcsin(\omega/\omega_{\rm max})$ and $\mu_k$ is the $k$-th Chebyshev moment
\begin{equation}
    \mu_k=\frac{1}{\cal N}\sum_{j=1}^{\cal N} U_k(\varepsilon_j).  \label{eq:chm}
\end{equation}

It is not possible to calculate the infinite number of the Chebyshev moments $\mu_k$, so we can cut off the series (\ref{eq:dec1}) and (\ref{eq:dec2}) at some $K$-th degree which is controlled by the desired accuracy of the calculations.  For the $\delta$-function it gives the following approximation
\begin{equation}
    \delta(\varepsilon-\varepsilon_j)\approx\frac{2}{\pi}\sqrt{1-\varepsilon^2}\sum\limits _{k=0}^K\gamma_k . U_k(\varepsilon_j)U_k(\varepsilon) \label{eq:dec4}
\end{equation}
The damping factors $\gamma_k$ were introduced to avoid Gibbs oscillations. With increasing $k$ these factors decrease gradually from 1 to 0 (for $k=K+1$). One of the best choice for $\gamma_k$ are Jackson damping factors~\cite{Weisse-2006}. The finite number of moments leads to the finite-width approximation of the $\delta$-function~\cite{Weisse-2006}
\begin{multline}
    \frac{2}{\pi}\sqrt{1-\varepsilon^2}\sum\limits _{k=0}^K\gamma_kU_k(\varepsilon_i)U_k(\varepsilon)\\
    \approx\frac{1}{\sqrt{2\pi\delta\varepsilon^2}}\exp\left[-\frac{(\varepsilon-\varepsilon_i)^2}{2\,\delta\varepsilon^2}\right].  \label{eq:decg}
\end{multline}
The width $\delta\varepsilon=\pi\sqrt{1-\varepsilon^2}/K$ corresponds to the frequency resolution $\delta\omega=\pi\sqrt{\omega_{\rm max}^2-\omega^2}/2K$. The greater the degree $K$ of the polynomial, the closer it to the delta function. As a result, we can approximately calculate the VDOS as
\begin{equation}
    g_\textsc{kpm}(\omega)=\frac{8\omega}{\pi\omega_{\rm max}^2}\sum_{k=0}^K\gamma_k\mu_k\sin((k+1)\varphi).   \label{eq:kpmdos}
\end{equation}
This sum can be calculated now by the Fast Fourier Transform (FFT) which is implemented in many mathematical libraries.

\begin{figure}[t]
    \centerline{\includegraphics[scale=0.38]{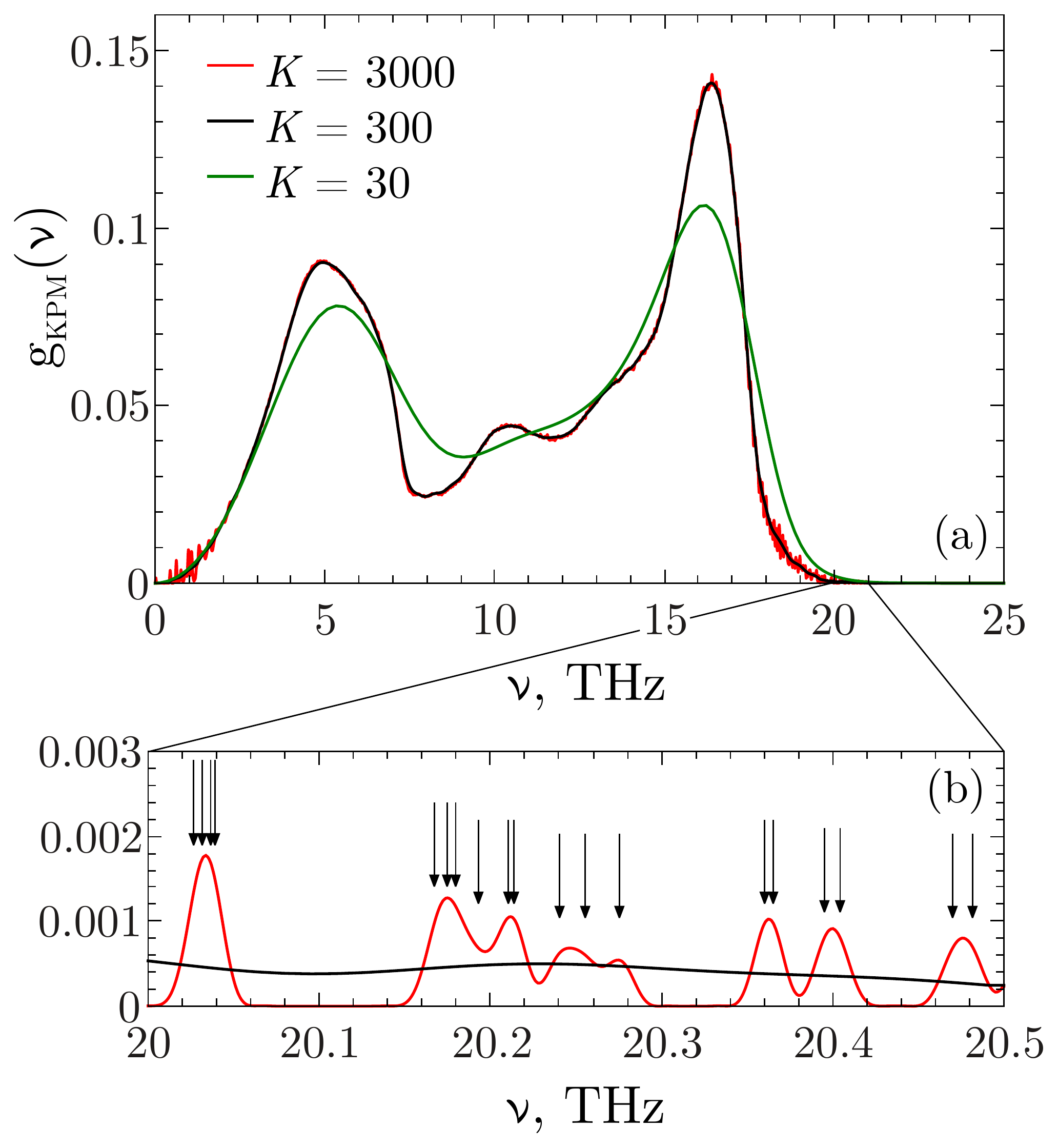}}
    \caption{(a) The calculated VDOS for $K=30$, $R=5$; $K=300$, $R=40$ and $K=3000$, $R=2000$. (b) The enlarged region 20--20.5 THz. Arrows show the exact positions of the eigenfrequencies.}
    \label{f.DOS-test}
\end{figure}

We turned the calculation of the VDOS $g(\omega)$ into calculation of Chebyshev moments $\mu_k$. Their definition (\ref{eq:chm}) can be written as
\begin{equation}
    \mu_k=\frac{1}{\cal N}\sum_{j=1}^{\cal N}\langle j|U_k(\tilde{M})|j\rangle  \label{eq:chsum}
\end{equation}
where we use ket notation $|j\rangle$ for the $j$th eigenvector of the matrix $\tilde{M}$ (the eigenvectors of the matrices $\tilde{M}$ and $M$ are the same). For a sufficiently large matrix $\tilde{M}$ the sum in Eq.~(\ref{eq:chsum}) can be replaced by the averaging over several realizations of a Gaussian random vector $|u_0\rangle$ with unit norm
\begin{equation}
    \mu_k=\overline{\langle u_0|U_k(\tilde{M})|u_0\rangle}.
\end{equation}

Indeed, let us expand  a random unit vector $|u_0\rangle$ over eigenvectors $|j\rangle$ of the matrix $\tilde{M}$
\begin{equation}
    |u_0\rangle=\sum\limits_j\beta_j|j\rangle,\quad\beta_j=\langle j|u_0\rangle.
\end{equation}
Therefore
\begin{equation}
    \langle u_0|U_k(\tilde{M})|u_0\rangle = \sum_{j=1}^{\cal N}|\beta_j|^2 U_k(\varepsilon_j).
\end{equation}
The random vector $|u_0\rangle$ is normalized, so $\overline{|\beta_j|^2}=1/{\cal N}$. As a result we have
\begin{equation}
    \overline{\langle u_0|U_k(\tilde{M})|u_0\rangle}=\frac{1}{\cal N}\sum_{j=1}^NU_k(\varepsilon_j)=\mu_k.
\end{equation}

Chebyshev moments $\mu_k$ for $k=0,\dots,K$ can be easily found by recurrence matrix-vector multiplications like (\ref{eq:Tk1})~--~(\ref{eq:Tk3})
\begin{align}
    |u_1\rangle &= 2\tilde{M}|u_0\rangle,   \label{eq:uk1} \\
    |u_k\rangle &= 2\tilde{M}|u_{k-1}\rangle-|u_{k-2}\rangle.   \label{eq:uk2}
\end{align}
It gives $|u_k\rangle = U_k(\tilde{M})|u_0\rangle$. At each step we calculate projection of $|u_k\rangle$ to the initial random vector $|u_0\rangle$
\begin{equation}
    m_k = \langle u_0|u_k\rangle.   \label{eq:muk}
\end{equation}
After averaging these projections over several number of realizations $R$ we obtain Chebyshev moments $\mu_k=\overline{m_k}$. Then the resulting VDOS is calculated making use of Eq.~(\ref{eq:kpmdos}).

Fig.~\ref{f.DOS-test} shows calculated VDOS for different number of moments $K$ taken into account. For a test purposes we use the dynamical matrix of our model of amorphous silicon with $N=32768$ atoms and parameter $\Lambda=21$. The number of realizations $R$ is big enough to neglect the statistical fluctuations (it is less then the linewidth in Fig.~\ref{f.DOS-test}(a)). We can see that $K=30$ is not enough because both peaks of the resulting VDOS are sufficiently broadened. On the other hand, the value of $K=3000$ is unnecessary big and we can see peaks from distinct eigenfrequencies (Fig.~\ref{f.DOS-test}(b)). We have found that the optimal value of $K$ is around $K=300$. For this value the KPM takes about one minute on a modern computer for calculation the VDOS.

Chebyshev polynomials of the first kind have similar recurrence relations (\ref{eq:Tk1})--(\ref{eq:Tk3}). So KPM can be implemented with the first kind polynomials as well as with the second kind polynomials. However, since usually $g(0)=0$, the second kind polynomials give better approximation in the low-frequency region.

\begin{figure}[t]
    \centerline{\includegraphics[scale=0.38]{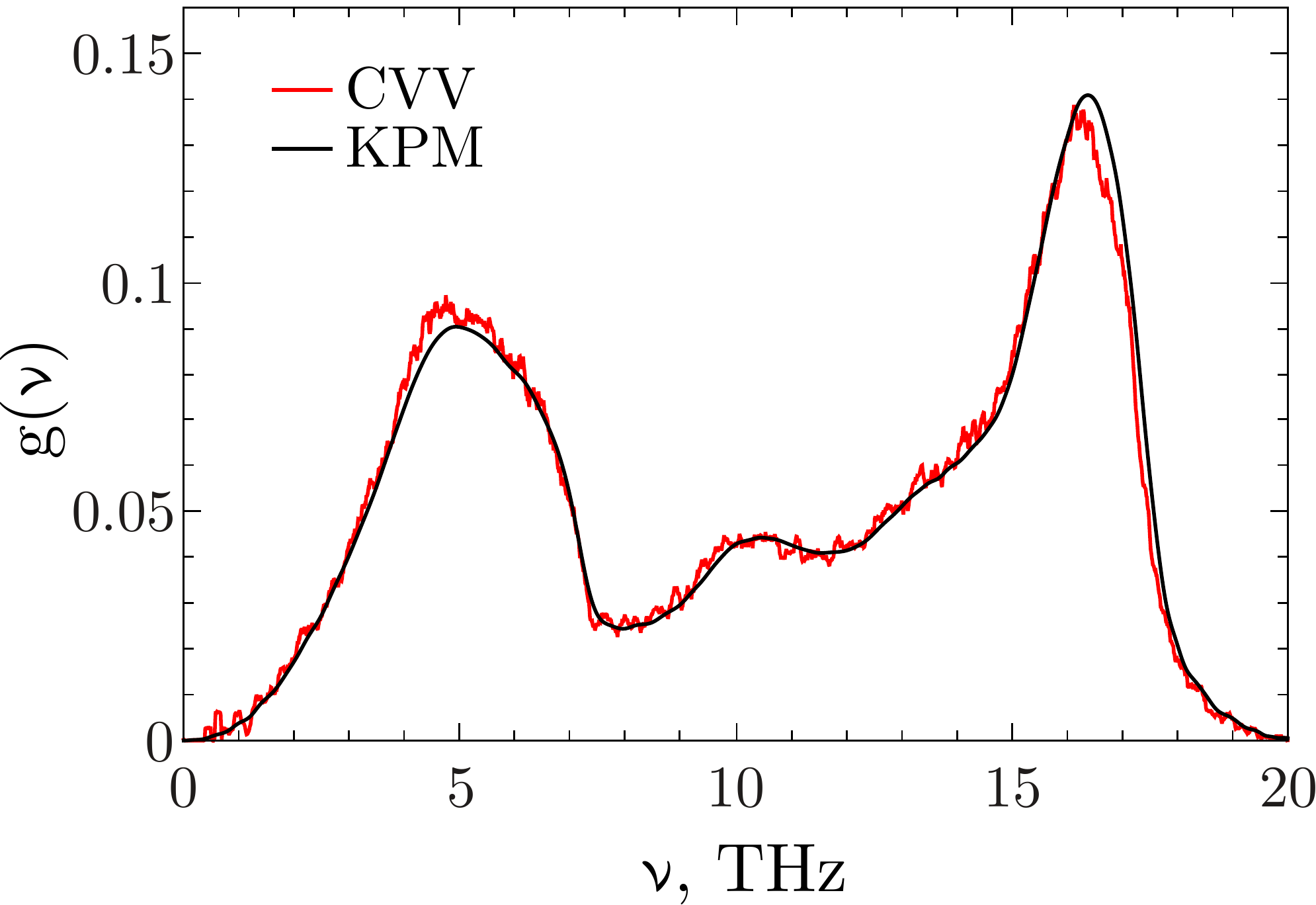}}
    \caption{A comparison of KPM and CVV method.}
    \label{f.DOS-CVV-KPM}
\end{figure}

We would like to emphasize here that the KPM has interesting physical meaning. The recurrence relations (\ref{eq:uk2}) reveals indeed a connection between KPM and CVV method. By definition, $\tilde{M} = I-2M/\omega_{\rm max}^2$ so
\begin{equation}
    |u_{k+1}\rangle = 2|u_k\rangle-|u_{k-1}\rangle-\delta t^2 M|u_k\rangle   \label{eq:uk}
\end{equation}
with $\delta t = 2/\omega_{\rm max}$ and $|u_k\rangle$ is the vector at step $k$ in the KPM. The Eq.~(\ref{eq:uk}) has the same form as the first step Verlet integration of equations for atomic displacements~\cite{Verlet-1967}
\begin{equation}
    {\bf u}_i(t+\delta t) = 2{\bf u}_i(t)-{\bf u}_i(t-\delta t)+\delta t^2 \ddot{\bf u}_i(t)   \label{eq:Verlet}
\end{equation}
where the acceleration $\ddot{\bf u}_i(t)$ of the $i$th atom is defined by the Newton's law (\ref{eq:motion}) and ${\bf u}_i(t)$ is the atomic displacement. Therefore, we can consider the integer variable $k$ as a discrete time $t=k\,\delta t$. Usually the time step $\delta t$ should be much less than $1/\omega_{\rm max}$ for reasonably small error in the integration procedure. The KPM relaxes this requirement to $\delta t = 2/\omega_{\rm max}$. The Chebyshev moments $\mu_k=\overline{\langle u_0|u_k\rangle}$ for $k=0,\ldots,K$ can be considered as auto-correlation functions of atomic displacements $\mu(t) = \overline{{\bf u}(0){\bf u}(t)}$ for $0\leq t\leq t_{\rm max} = K\,\delta t$. The resulting VDOS (\ref{eq:kpmdos}) is the Fourier transform of the Chebyshev moments. The finite frequency resolution $\delta \omega$ corresponds to the finite-time limit $1/t_{\rm max}$. Thus the KPM is similar to the CVV method (\ref{eq:cvvdos}), however the remarkable properties of the Chebyshev polynomials allows to take a big time step $\delta t = 2/\omega_{\rm max}$ instead of much smaller time step $\delta t\ll 1/\omega_{\rm max}$.

\section{Eigenvector analysis by KPM\label{sec:evecKPM}}

The correlation function (Sec.~\ref{sec:modes}) and the dynamical structure factor (Sec.~\ref{sec:sf}) are two of the main eigenvector characteristics. The former  shows the spatial correlations of vibrations and the latter shows the structure in the reciprocal space. The direct eigenvector analysis with full diagonalization of the dynamical matrix takes too much computational time. In this section we show how to properly modify the KPM (Appendix~\ref{sec:KPM}) for eigenvector analysis.

The definition of the correlation function (\ref{eq:corr}) is similar to the definition of VDOS (\ref{eq:prDOS}), but it contains in addition (as multiplier) the bilinear form of the eigenmodes, which depends on the external parameter ${\bf r}$. In this case we can the whole evaluation procedure of the KPM (Appendix \ref{sec:KPM}) with modified projection (\ref{eq:muk}) of the state $|u_k\rangle$ to the initial state $|u_0\rangle$
\begin{equation}
    m_k({\bf r}) = \langle {\bf u}_0({\bf r}+{\bf r}')\cdot{\bf u}_k({\bf r}')\rangle_{{\bf r}'}.
\end{equation}
Thus, for each fixed ${\bf r}$ we can efficiently calculate the correlation function as a sum of Chebyshev polynomials.

The same idea is applicable to the dynamical structure factor. It depends on the spatial Fourier transform of the eigenmodes $F_\eta({\bf q}, \omega)$, which also contains bilinear form of the eigenmodes. Therefore we can use modified projections (\ref{eq:muk}), which are slightly different for longitudinal and transverse components
\begin{align}
    m_k^L({\bf q}) &= {\cal N}\biggl(\sum_{i=1}^N\hat{\bf q}\cdot{\bf u}^0_ie^{i{\bf q}{\bf R}_i}\biggr)^*\biggl(\sum_{i=1}^N\hat{\bf q}\cdot{\bf u}^k_ie^{i{\bf q}{\bf R}_i}\biggr),\\
    m_k^T({\bf q}) &= {\cal N}\biggl(\sum_{i=1}^N\hat{\bf q}\times{\bf u}^0_ie^{i{\bf q}{\bf R}_i}\biggr)^*\!\cdot\biggl(\sum_{i=1}^N\hat{\bf q}\times{\bf u}^k_ie^{i{\bf q}{\bf R}_i}\biggr).
\end{align}

A more general and detailed information about the calculation of eigenvector characteristics and Green's function can be found in the review~\cite{Weisse-2006}.

\section{A general decomposition to transverse and longitudinal vibrations\label{sec:LT}}

Low-frequency vibrations below Ioffe-Regel criterion are well-defined plane waves (see Section~\ref{sec:sf} for details). In continuous medium approximation the displacement fields ${\bf u}({\bf r})$ for longitudinal (L) and transverse (T) waves have a form
\begin{gather}
    {\bf u}_\eta({\bf r}) = {\bf u}_\eta^{(0)}\exp(i{\bf qr}), \quad \eta=L,T \\
    {\bf u}_L^{(0)} \parallel {\bf q}, \quad {\bf u}_T^{(0)} \perp {\bf q}.  \label{eq:planeLT}
\end{gather}
However, above the Ioffe-Regel criterion the wave vector ${\bf q}$ is ill-defined so we cannot use the definition (\ref{eq:planeLT}) in a general case.

The transverse displacement field ${\bf u}_T({\bf r})$ has zero divergence, therefore it conserves the local volume. A natural analog of the local volumes in amorphous media are Voronoi cells constructed around each atom. By definition the Voronoi cell ${\cal V}_i$, associated with atom $i$ is the set of all points in the surrounding space whose distance to the atom $i$ is not greater than their distances to the other atoms $j$~\cite{Voronoi-1991}.

Displacements of atoms ${\bf u}_i$ in amorphous media may (or may not) change volumes of Voronoi cells. We will call the displacement of atoms ${\bf u}_i$ to be transverse if it does not change the volumes of all Voronoi cells. For that let us introduce a matrix $A$ which is responsible for the relative change of the $i$th Voronoi cell volume $V_i$ under $j$th atom displacement in the direction $\alpha$
\begin{equation}
    A_{i,j\alpha} = \frac{1}{V_i}\frac{\partial V_i}{\partial r_{j\alpha}}.   \label{eq:A}
\end{equation}
The explicit formula for the matrix $A$ can be derived from geometry only~\cite{Beltukov-2015}. Using this matrix the displacement of $j$th atom in the direction $\alpha$,  $u_{j\alpha}$ results in the following relative change of the Voronoi cell volumes $A_{i,j\alpha}u_{j\alpha}$. Summing over all $j$ and $\alpha$ gives the relative change of the $i$th Voronoi cell volume
\begin{equation}
    \varepsilon_i = \sum_{j\alpha}A_{i,j\alpha}u_{j\alpha}.
\end{equation}
In the bra-ket notation this equation reads $|\varepsilon\rangle = A|u\rangle$ where $A$ is a rectangular $N\times3N$ matrix (with $N$ being the number of atoms) and $|u\rangle$ is a displacement vector with $3N$ elements. The matrix $A$ is a discrete analog of the divergence operator.

By definition the transverse component $|u_T\rangle$ of the arbitrary $|u\rangle$ satisfies to equation  $A|u_T\rangle=0$, i.e. $|u_T\rangle$ is the projection of the displacement $|u\rangle$ to the \textit{null space} of the matrix $A$. The longitudinal component $|u_L\rangle$ is a remaining orthogonal component of the displacement field and it is the projection of $|u\rangle$ to the \textit{row space} of the matrix $A$. These projections have the following forms~\cite{Meyer-2000}
\begin{equation}
    |u_\eta\rangle = P_\eta |u\rangle,
\end{equation}
where
\begin{align}
    P_L &= A^T(AA^T)^{-1}A,\\
    P_T &= I-A^T(AA^T)^{-1}A.
\end{align}
One can easy check that $A |u_T\rangle = 0$ and $\langle u_L| u_T\rangle = 0$.

Thus $P_L|j\rangle$ and $P_T|j\rangle$ are projections of the eigenmode $|j\rangle$ to longitudinal and transverse components respectively. Therefore, the total VDOS $g(\omega)$ can be decomposed into the longitudinal and transverse components in general case independently on frequency $\omega$
\begin{gather}
    g(\omega) = \frac{1}{\cal N}\sum_{j=1}^{\cal N} \delta(\omega-\omega_j) = g_L(\omega) + g_T(\omega), \\
    g_\eta(\omega) = \frac{1}{\cal N}\sum_{j=1}^{\cal N} \langle j|P_\eta|j\rangle\delta(\omega-\omega_j), \quad\eta = L,T  \label{eq:DOSLT}
\end{gather}
where eigenfrequency $\omega_j$ corresponds to the eigenvector $|j\rangle$. In three dimensions we have $\int g_L(\omega)d\omega=1/3$ and  $\int g_T(\omega)d\omega=2/3$.

The definition of longitudinal and transverse components of the VDOS (\ref{eq:DOSLT}) contains the bilinear form of the eigenmode (as well as the correlation function (\ref{eq:corr}) and the Fourier transform (\ref{eq:Fl}), (\ref{eq:Ft})). Therefore one can apply the KPM with the modified projection (\ref{eq:muk})
\begin{equation}
   m_k^\eta = \langle u_0|P_\eta|u_k\rangle.   \label{eq:muwk}
\end{equation}

The results of this method are discussed in the Sec.~\ref{sec:DOS}.


\begin{thebibliography}{999}


\bibitem{aSi-Appli} J. Zhu {\it et al.}, Nano Lett. {\bf 9}, 279 (2009).

\bibitem{b.feldman0} J. L. Feldman, M. D. Klug\'e, P. B. Allen and F. Wooten, Phys. Rev. B {\bf 48}, 12589 (1993).
\bibitem{b.feldman1} J. L. Feldman, P. B. Allen and S. R. Bickham, Phys. Rev. B {\bf 59}, 3551 (1999).
\bibitem{b.allen} P. B. Allen, J. L. Feldman, J. Fabian and F. Wooten, Phil. Mag. B {\bf 79}, 1715 (1999).

\bibitem{Tg-aSi2004} A. Hedler, S. L. Klaumunzer and W. Wesch, Nat. Mater. {\bf 3}, 804 (2004).
\bibitem{Tg-aSi2003} S. Sastry and C. A. Angell, Nat. Mater. {\bf 2}, 739 (2003).
\bibitem{Tg-aSi2004b} P. F. McMillan, Nat. Mater. {\bf 3}, 755 (2004).

\bibitem{Wyart2005} M. Wyart, S. R. Nagel and T. A. Witten Europhysics Letters {\bf 72}, 486 (2005).
\bibitem{Buchenau1984} U. Buchenau, N. Nucker and A. J. Dianoux, Phys. Rev. Lett. {\bf 53}, 2316 (1984).
\bibitem{Buchenau1986} U. Buchenau, M. Prager, N. Nucker, A. J. Dianoux, N. Ahmad and W. A. Phillips, Phys. Rev. B {\bf 34}, 5665 (1986).
\bibitem{Ruocco2013} A. Marruzzo, W. Schirmacher, A. Fratalocchi and G. Ruocco, Scientific Reports {\bf 3}, 1407 (2013).
\bibitem{giordano_2010} V. M. Giordano and G. Monaco, PNAS {\bf 107}, 21985 (2010).
\bibitem{giordano_2011} G. Baldi, V. M. Giordano and G. Monaco, Phys. Rev. B {\bf 83}, 174203 (2011).
\bibitem{Parshin1992} U. Buchenau, Yu. M. Galperin, V. L. Gurevich, D. A. Parshin, M. A. Ramos and H. R. Schober, Phys. Rev. B {\bf 46}, 2798 (1992).
\bibitem{Parshin1993} V. I. Gurevich, D. A. Parshin, J. Pelous and H. R. Schober, Phys. Rev. B {\bf 48}, 16318 (1993).
\bibitem{Parshin2003} V. I. Gurevich, D. A. Parshin and H. R. Schober, Phys. Rev. B {\bf 67}, 094203 (2003).
\bibitem{Parshin1994} D. A. Parshin, Phys. Rev. B {\bf 49}, 9400 (1994).
\bibitem{Sokolov1992} A. P. Sokolov, A. Kisliuk, M. Soltwisch and D. Quitmann, Phys. Rev. Lett. {\bf 69}, 1540 (1992).
\bibitem{Sokolov1986} V. K. Malinovsky and A. P. Sokolov, Solid State Commun. {\bf 57}, 757 (1986).
\bibitem{b.schirmacher1} W. Schirmacher, G. Diezemann and C. Ganter, Phys. Rev. Lett. {\bf 81}, 136 (1998).
\bibitem{b.schirmacher2} W. Schirmacher, C. Tomaras, B. Schmid, G. Baldi, G. Viliani, G. Ruocco and T. Scopigno, Cond. Matt. Phys. {\bf 13}, 23606 (2010).

\bibitem{Wyart2010b} M. Wyart, Europhysics Letters {\bf 84}, 64001 (2010).
\bibitem{Parshin2007-Review} D. A. Parshin, H. R. Schober and V. L. Gurevich, Phys. Rev. B {\bf 76}, 064206 (2007).
\bibitem{Pohl1971} R. C. Zeller and R. O. Pohl, Phys. Rev. B {\bf 4}, 2029 (1971).
\bibitem{b.malinovsky} V. K. Malinovsky, V. N. Novikov, P. P. Parshin, A. P. Sokolov and M. G. Zemlyanov, Europhys. Lett. {\bf 11}, 43 (1990).


\bibitem{Sokolov1991} A. P. Sokolov, A. P. Shebanin, O. A. Golikova and M. M. Mesdrogina, J. Phys.: Condens. Mat. {\bf 3}, 9887 (1991).
\bibitem{aSi-VDOS-Exp} W. A. Kamitakahara, C. M. Soukoulis, H. R. Shanks, U. Buchenau and G. S. Grest, Phys. Rev. B {\bf 36}, 6539 (1987).
\bibitem{Thorpe1973} M. F. Thorpe, Phys. Rev. B {\bf 8}, 5352 (1973).
\bibitem{He} Y. He, D. Donadio and G. Galli, Appl. Phys. Lett. {\bf 98}, 144101 (2011).
\bibitem{b.fabian2} J. Fabian, J. L. Feldman, C. S. Hellberg and S. M. Nakhmanson, Phys. Rev. B {\bf 67}, 224302 (2003).
\bibitem{b.finkemeier} F. Finkemeier and W. von Niessen, Phys. Rev. B {\bf 63}, 235204 (2001).
\bibitem{b.resp-finkemeier} S. M. Nakhmanson, D. A. Drabold and N. Mousseau, Phys. Rev. B {\bf 66}, 087201 (2002).

\bibitem{b.christie} J. K. Christie, S. N. Taraskin and S. R. Elliott, J. Non-Cryst. Solids {\bf 353}, 2272 (2007).
\bibitem{b.marinov} M. Marin\o v and N. Zotov, Phys. Rev. B {\bf 55}, 2938 (1997).

\bibitem{Duval2007} E. Duval, A. Mermet, and L. Saviot, Phys. Rev. B {\bf 75}, 024201 (2007).

\bibitem{Phillips-book} W. A. Phillips, Amorphous Solids Low-Temperature Properties, Springer ed., Berlin, 1981.
\bibitem{taraskin-1997} S. N. Taraskin and S. R. Elliott, Phys. Rev. B {\bf 56}, 8605 (1997).
\bibitem{Kittel-book} C. Kittel. {\it Introduction to solid state physics} (Wiley, 2005).

\bibitem{taraskin2000} S. N. Taraskin and S. R. Elliott, Phys. Rev. B {\bf 61}, 12017 (2000).
\bibitem{b.tanguy2010} A. Tanguy, B. Mantisi and M. Tsamados, Europhys. Lett. {\bf 90} 16004 (2010).

\bibitem{Lemaitre2004} C. E. Maloney and A. Lemaitre, Phys. Rev. Lett. {\bf 93}, 195501 (2004).
\bibitem{Lemaitre2006} C. E. Maloney and A. Lemaitre, Phys. Rev. E {\bf 74}, 016118 (2006).
\bibitem{b.tanguy20061} A. Tanguy, F. Leonforte and J.-L. Barrat, Eur. Phys. J. E {\bf 20}, 355 (2006).

\bibitem{Ioffe} A. F. Ioffe, A. R. Regel, Prog. Semicond. 4, 237 (1960).
\bibitem{taraskin2000b} S. N. Taraskin and S. R. Elliott, Phys. Rev. B 61, 12031 (2000).
\bibitem{schober2004} H R Schober, J. Phys.: Condens. Matter, 16, S2659 (2004).

\bibitem{Beltukov2013} Y. M. Beltukov, V. I. Kozub and D. A. Parshin, Phys. Rev. B {\bf 87}, 134203 (2013).

\bibitem{Damart2015} T. Damart, V. Giordano and A. Tanguy, Phys. Rev. B {\bf 92}, 094201 (2015).

\bibitem{taraskin2002}  S. N. Taraskin and S. R. Elliott, J. Phys.:Condens. Matter {\bf 14}, 3143 (2002).

\bibitem{larkin} J. M. Larkin and A. J. H. McGaughey, Phys. Rev. B {\bf 89}, 144303 (2014).
\bibitem{b.tanguy2002} A. Tanguy, J. P. Wittmer, F. Leonforte and J.-L. Barrat, Phys. Rev. B, {\bf 66} 174205-1-17 (2002).
\bibitem{Wyart2010} V. Vitelli, N. Xu, M. Wyart, A. J. Liu and S. R. Nagel, Phys. Rev. E {\bf 81}, 021301 (2010).
\bibitem{Barrat2014} H. Mizuno, S. Mossa and J.-L. Barrat, PNAS {\bf 111}, 11949. (2014).

\bibitem{b.taraskin} S. N. Taraskin, Y. L. Loh, G. Natarajan and S. R. Elliott, Phys. Rev. Lett. {\bf 86}, 1255 (2001).


\bibitem{b.stillinger} F. H. Stillinger, T. A. Weber, Phys. Rev. B {\bf 31}, 5262 (1985).

\bibitem{b.fusco2010} C. Fusco, T. Albaret and A. Tanguy, Phys. Rev. E {\bf 82} 066116 (2010).
\bibitem{b.fusco2014} C. Fusco, T. Albaret and A. Tanguy, Eur. Phys. J. E {\bf 37}, 43 (2014).
\bibitem{b.pizza2013} L. Pizzagalli, J. Godet, J. Guenole, S. Brochard, E. Holmstrom, K. Nordlund and T. Albaret, J. Phys.: Condens. Mat. {\bf 25}, 055801 (2013).



\bibitem{b.lammps} S. J. Plimpton, J. Comput. Phys. {\bf 117} (1995). See also http://lammps.sandia.gov.



\bibitem{feast2009} E. Polizzi, Phys. Rev. B {\bf 79}, 115112 (2009).

\bibitem{b.tanguy20062} F. L\'eonforte, A. Tanguy, J. P. Wittmer and J.-L. Barrat, Phys. Rev. Lett. {\bf 97} 055501 (2006).

\bibitem{Tubino-1972} R. Tubino, L. Piseri and G. Zerbi, J. Chem. Phys. {\bf 56}, 1022 (1972).



\bibitem{b.schober} H. R. Schober and C. Oligschleger, Phys. Rev. B {\bf 53}, 11469 (1996).

\bibitem{Beltukov2011} Y. M. Beltukov and D. A. Parshin, Phys. Solid State 53, 151 (2011) [Fizika Tverdogo Tela 53, 142 (2011)].

\bibitem{vanTiggelen} A. Lagendijk, B. A. van Tiggelen, Physics Reports {\bf 270}, 143 (1996).

\bibitem{Page2009} S. Faez, A. Strybulevych, J. H. Page, A. Lagendijk and B. van Tiggelen, Phys. Rev. Lett. {\bf 109}, 155703 (2009).
\bibitem{Localization} C. Castellani and L. Peliti, J. Phys. A 19, L429 (1986).



\bibitem{shintani} H. Shintani and H. Tanaka, Nat. Mat. {\bf 7}, 870 (2008).

\bibitem{Ruffle2006} B. Ruffle, G. Guimbretiere, E. Courtens, R. Vacher and G. Monaco, Physical Review Letters {\bf 96}, 045502 (2006).





\bibitem{taraskin2003} J. J. Ludlam, S. N. Taraskin and S. R. Elliott, Phys. Rev. B {\bf 67}, 132203 (2003).
\bibitem{taraskin2005} J. J. Ludlam, S. N. Taraskin, S. R. Elliott and D. A. Drabold, J. Phys.: Condens. Matter {\bf 17}, L321 (2005).


\bibitem{Dove} M. T. Dove {\it Introduction to Lattice Dynamics} (Cambridge University Press, Cambridge, 1993).

\bibitem{Sampoli-1998} M. Sampoli et al. Phil. Mag. B, {\bf 77}, 473 (1998).

\bibitem{Weisse-2006} A. Wei\ss e et al. Rev. Mod. Phys. {\bf 78}, 275 (2006).
\bibitem{Silver-1997} R. N. Silver and H. R\:oder. Phys. Rev. E {\bf 56}, 4822 (1997).

\bibitem{Verlet-1967} L. Verlet, Phys. Rev. {\bf 159}, 98 (1967).


\bibitem{Voronoi-1991} F. Aurenhammer, ACM Computing Surveys {\bf 23}, 345 (1991).
\bibitem{Beltukov-2015} Y. M. Beltukov, C. Fusco, A. Tanguy, D. A. Parshin,  J. Phys.: Conf. Ser. (to be published, arXiv:1508.04252).

\bibitem{Meyer-2000} C. D. Meyer, Matrix Analysis and Applied Linear Algebra, SIAM (2000).


\end{thebibliography}
\end{document}